\documentclass[aoas,MSNbibl,nameyear,rotating,seceqn,noautosecdot,dvips]{arximspdf}
\usepackage{dcolumn}
\usepackage{graphicx}

\doi{10.1214/13-AOAS648} %kopijuoti is PTS
\volume{7}
\issue{3}
\pubyear{2013}
\firstpage{1612}
\lastpage{1639}

\makeatletter
\newcolumntype{d}[1]{D{.}{.}{#1}}
\newcommand{\x}{\mathbf{x}}
\newcommand{\Sep}{ | }
\newcommand{\pkg}[1]{{\textit{#1}}}
\newcommand{\code}[1]{\texttt{#1}}

\setattribute{abstract}    {skip} {20\p@}
\setattribute{keyword}     {skip} {8\p@}
\setattribute{frontmatter} {skip} {0\p@ plus 3\p@ minus 3\p@}

\makeatother

\begin{document}
\begin{frontmatter}

\title{Influencing elections with statistics: Targeting voters with
logistic regression trees}
\runtitle{Targeting voters with logistic regression trees}

\begin{aug}
\author[a]{\fnms{Thomas} \snm{Rusch}\corref{}\ead[label=e1]{thomas.rusch@wu.ac.at}\thanksref{t1}},
\author[b]{\fnms{Ilro} \snm{Lee}\ead[label=e2]{ilro.lee@unsw.edu.au}\thanksref{t2}},
\author[c]{\fnms{Kurt} \snm{Hornik}\ead[label=e3]{kurt.hornik@wu.ac.at}\thanksref{t1}},
\author[d]{\fnms{Wolfgang}~\snm{Jank}\ead[label=e4]{wjank@usf.edu}\thanksref{t3}}
\and
\author[e]{\fnms{Achim} \snm{Zeileis}\ead[label=e5]{Achim.Zeileis@R-project.org}\thanksref{t4}}
\runauthor{T. Rusch et al.}
\affiliation{WU (Vienna University of Economics and Business)\thanksmark{t1},
University of New South Wales\thanksmark{t2},
University of South Florida\thanksmark{t3} and
Universit\"at Innsbruck\thanksmark{t4}}
\address[a]{T. Rusch\\
Center for Empirical Research Methods\\
WU (Vienna University of Economics and Business)\\
Augasse 2-6\\
1090 Vienna\\
Austria\\
\printead{e1}}

\address[b]{I. Lee\\
School of Management\\
Australian School of Business\\
University of New South Wales\hspace*{18pt}\\
Sydney NSW 2052\\
Australia\\
\printead{e2}}

\address[c]{K. Hornik\\
Institute for Statistics and Mathematics\\
Department of Finance, Accounting and Statistics\\
WU (Vienna University of Economics and Business)\\
Augasse 2-6\\
1090 Vienna\\
Austria\\
\printead{e3}}

\address[d]{W. Jank\\
Department of Information Systems \\
\quad and Decision Sciences\\
College of Business\\
University of South Florida\\
4202 E. Fowler Ave., BSN 3403\\
Tampa, Florida 33620-5500\\
USA\\
\printead{e4}}

\address[e]{A. Zeileis\\
Department of Statistics\\
Faculty of Economics and Statistics\\
Universit\"at Innsbruck\\
Universit\"atsstr.~15\\
6020 Innsbruck\\
Austria\\
\printead{e5}}
\end{aug}

% HISTORY:
\received{\smonth{3} \syear{2012}}
\revised{\smonth{3} \syear{2013}}

% ABSTRACT
%
\begin{abstract}
In political campaigning substantial resources are spent on voter
mobilization, that is, on identifying and influencing as many people as
possible to vote. Campaigns use statistical tools for deciding whom to
target (``microtargeting''). In this paper we describe a nonpartisan
campaign that aims at increasing overall turnout using the example of
the 2004 US presidential election. Based on a real data set of
19,634 eligible voters from Ohio, we introduce
a modern statistical framework well suited for carrying out the main
tasks of voter targeting in a single sweep: predicting an individual's
turnout (or support) likelihood for a particular cause, party or
candidate as well as data-driven voter segmentation. Our framework,
which we refer to as LORET (for LOgistic {RE}gression {T}rees), contains standard methods such as
logistic regression and classification trees as special cases and
allows for a synthesis of both techniques. For our case study, we
explore various LORET models with different regressors in the logistic
model components and different partitioning variables in the tree
components; we analyze them in terms of their predictive accuracy and
compare the effect of using the full set of available variables against
using only a limited amount of information. We find that augmenting a
standard set of variables (such as age and voting history) with
additional predictor variables (such as the household composition in
terms of party affiliation) clearly improves predictive accuracy. We
also find that LORET models based on tree induction beat the
unpartitioned models. Furthermore, we illustrate how voter segmentation
arises from our framework and discuss the resulting profiles from a
targeting point of view.
\end{abstract}

% KEYWORDS

\end{frontmatter}
\newpage

%s1 #&#
\section{\texorpdfstring{Introduction.}{Introduction}}
``Decisions are made by those who show up,'' said President Bartlet,
a character from a popular TV show, \textit{The West Wing}. The character
in the show used the line to motivate a college audience to voice their
opinion by showing up at the polls. Getting eligible voters to actually
vote (``get-out-the-vote;'' GOTV) is an important goal in countries
with a democratic political system and a lot of resources are spent on
achieving that goal. Take the 2012 US presidential race, for example.
In that year, the world witnessed the amount of money raised and spent
by the campaigns reaching unprecedented heights. By spending over
USD~$1.5$~billion, the Obama and Romney campaigns tried to mobilize
eligible voters to engage in the political process and cast their vote
on November 6th.

%s1.1 #&#
\subsection{\texorpdfstring{Campaigning, mobilization and turnout.}{Campaigning, mobilization and turnout}}
The impact of partisan campaigning or nonpartisan get-out-the-vote
efforts on mobilization and turnout has been subject to numerous
scientific investigations over the last 20 years. Examples include
\citeauthor{whitelock2010}'s (\citeyear{whitelock2010}) survey on the effect of campaigning and turnout in
the UK and Germany or \citet{karp2007} who surveyed the relationship
between party contacts and turnout in 23 countries (old and new
democracies). See also \citet{holbrook2005} for an overview of recent
studies. Starting from an early ``minimal effect'' hypothesis [\citet
{finkel1993}, i.e., the idea that political campaigns barely influence
turnout], there is evidence in the literature that campaigning does
indeed have a measurable effect on persuasion or mobilization of the
electorate [\citet{holbrook2005}], which is supported by a number of
experimental studies, for example, \citet
{nickerson2006}, \citeauthor{gerber2000a} (\citeyear{gerber2000a,gerber2000b}),
\citet{green2003a,philips2008,hansen2009,arceneaux2009}.\footnote{Although the literature seems to have not yet
reached a consensus, especially with respect to partisan GOTV; see
\citet
{cardy2005,gerber2003,panagopoulos2009b}.}

Reinforced by these results, campaigns are spending large amounts of
money on mobilizing voters. However, one cannot simply equate higher
spending with higher turnout. Take the United States, for example,
where the ``professionalization'' [\citet{muller1999}] of campaigning
had its origin [\citet{plasser2000}] and spread to many democratic
countries all over the world [\citet{sussman2003}]. Arguably, nowhere
else is political campaigning a bigger business then in the US and
nowhere else is more money being spent on convincing people to cast
their ballots. Despite increased political consultancy, monumental
campaign efforts and large out-laying of resources, the average voter
turnout since 1980 during the Presidential election years has only been
$56\%$; see also Table~\ref{tblvoterturnout}. This raises questions
about the effectiveness of campaigns' voter mobilization strategies.

%
%t1 #&#
\begin{table}
\caption{Individual and aggregated turnout rate (votes for highest
office divided by the voting-eligible population) for presidential
elections in the United States and the money spent by all candidates
(in million USD). The fourth column lists the real expenditures
(inflation-adjusted at 2008 rates). Sources: \citet{mcdonald2012} and
\protect\url{http://www.opensecrets.org/}, accessed 11-21-2012 (all elections
until 2008), \citet{Wikielection} and \textit{The New York Times}
(\protect\url{http://elections.nytimes.com/2012/campaign-finance}, accessed
11-21-2012) (2012 election). Inflation-adjustment has been done with
\protect\url{http://www.bls.gov/data/inflation_calculator.htm} at 11-21-2012}
\label{tblvoterturnout}
\begin{tabular*}{\textwidth}{@{\extracolsep{\fill}}ld{2.1}d{4.1}d{4.1}@{}}
\hline
& & \multicolumn{1}{c}{\textbf{Expenditures}} & \multicolumn
{1}{c@{}}{\textup{\textbf{Real
expenditures}}} \\
\multicolumn{1}{@{}l}{\textbf{Year}} & \multicolumn{1}{c}{\textbf
{Turnout (in \%)}} & \multicolumn{1}{c}{\textbf{(in mill.~USD)}} &
\multicolumn{1}{c@{}}{\textbf{(at 2008 rates)}} \\
\hline
2012 & 57.5 & 1605.2& 1494.7\\
2008 & 61.7 & 1324.7& 1324.7\\
2004 & 60.1 & 717.9& 818.2\\
2000 & 54.2 & 343.1 & 429.0\\
1996 & 51.7 & 239.9 & 329.2\\
1992 & 58.1 & 192.2 & 295.0\\
1988 & 52.8 & 210.7 & 383.5\\
1984 & 55.2 & 103.6 & 214.7\\
1980 & 54.2 & 92.3 & 241.2 \\[3pt]
Mean & 56.2 & 536.6 & 614.5 \\
Sd & 3.4 & 562.5 & 486.2 \\
Min & 51.7 & 92.3 & 214.7 \\
Max & 61.7 & 1605.7 & 1494.7 \\
\hline
\end{tabular*}
\end{table}

%
%f1 #&#
\begin{figure}

\includegraphics{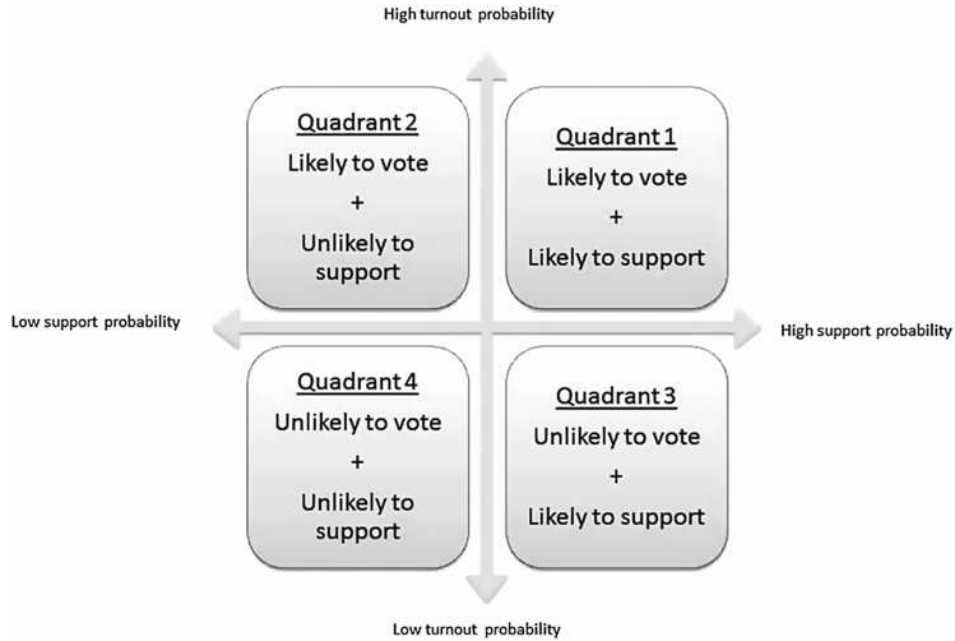}

\caption{The usual partisan campaign classification of targeting
groups.}\label{figquadrants}
\end{figure}

%s1.2 #&#
\subsection{\texorpdfstring{How is targeting carried out?}{How is targeting carried out?}}
Voter mobilization is a two-step process [cf. \citet{goldstein2002b}].
In the first step, campaigns need to identify people suitable to direct
their mobilization efforts at (also known as voter targeting). The
second step involves crafting measures that best motivate these people
to turn up at the polls, that is, to assure the effectiveness of
mobilization. The latter step includes decisions on which tactics best
translate to mobilization and has been investigated by researchers in
the political and social sciences or marketing [for an overview of
which measures to use see, e.g., \citet{green2008}]. The first step
(\emph{identifying} the ``right'' recipients for mobilization messages)
has, to the best of our knowledge, been addressed rather infrequently
in the scientific literature. Notable exceptions are \citet
{wielhouwer2003,parry2008,murray2010} or \citet{imai2011}.

When identifying people to target, campaigns typically first assess two
important aspects for each eligible voter: (a) likelihood of support
(for a particular cause, party or candidate) and (b) likelihood to
turnout at the polls [\citet{malchow08,Issenberg2012b}]. Using these
two assessments, each voter can be schematically classified into one of
four possible categories (or ``quadrants,'' see Figure~\ref{figquadrants}). For voters that are classified into quadrant 1
(likely to vote and likely to support), campaigns usually allocate few
resources on mobilization (but these voters might be asked to help out
with the campaign). Eligible voters assigned to quadrant 2 (likely to
vote but unlikely to support) are ``targeted for support'' by the
campaigns, as they can be persuaded to become supporters. In quadrant 3
(unlikely to vote but likely to support), the focus of the targeting
effort will be on mobilization for turnout (``targeting for turnout'').
For both quadrants 2 and 3 the campaigns will use targeted messages.
The messages could be customized to individuals based on their
demographic and behavioral data (``voter profiles'').
Voters classified to belong to quadrant 4 (unlikely to vote and
unlikely to support) will typically not be targeted by a campaign [see
also \citet{Issenberg2012b}].

In order to populate quadrants 1--4, campaigns need rich voter data and
powerful data models that can predict, for each individual person,
his/her probability of support or turnout with as high accuracy as
possible. In some countries, voting data that can be used to explain
and predict voting behavior is available as public data. In the US, for
example, states collect and report voter registration information and
make them publicly available. Collection is done at the county level
and the data are only available in aggregated fashion. Individual
voting data is usually not readily and easily accessible [\citet
{congress}]. Data for targeting also arrives in the form of proprietary
information, offered by data vendors who supply individual-level data
and add considerable details about voter behavior and demographics. In
many countries, proprietary sources from market research companies are
the only way to obtain data for targeting, as public data are scarce.

For all data sources the most important predictor variables typically
collected are records of the (individual) voting history. The ability
of voting history as a predictor for future election attendance has
long been recognized [e.g., \citet{denny2009}] and, consequently, for
targeting purposes voting history is heavily relied on [\citet
{goldstein2002b,malchow08}]. Additional predictive power has been
found in sociodemographic or personality variables like age, income and
party affiliation.

While campaigns can collect an abundance of predictor variables with
ease, collecting information on the target variable poses a more
challenging problem. Supervised classification methods require a known
target (i.e., observations on the response variable) in order to train
the model. In the case of an election, the target (i.e., whether a
person will truly turnout or support) is not known until the election
is over. Campaigns therefore have to rely on suitable proxy target
variables which should most accurately resemble the true outcome. The
usage of proxies renders the application of supervised classification
procedures during or before the election feasible. While many proxies
(e.g., an earlier election) are imaginable and the choice may vary
between campaigns, proxy variables often arrive in the form of
carefully designed polls about voting intention. For example, the Obama
2012 campaign conducted short, parallel survey polls on random samples
of 8000 to 9000 voters from ``battleground states'' every night during
the final phase of the campaign [\citet{Blumenthal2012}]. For the rest
of this paper we only consider the situation of either employing the
true outcome or proxy variables derived from surveys, but we have also
investigated the use of proxy variables derived from previous election
outcomes; see the supplementary material [\citet{supplementb}].

Campaigns often have access to similar sources of information, but the
way the information is processed, modeled and ultimately acted on can
be very diverse. Traditionally, campaigns have relied on simple
deterministic rules for choosing whom to target by, for example, using
information from the last four comparable elections as the main
predictors for future voting behavior. Intuitively, someone who voted
in all four out of the last four elections is seen as a likely voter,
whereas someone who did not vote in any of the four elections is
considered unlikely to vote in the upcoming election. However,
predicting the behavior of a person with a mixed voting pattern (i.e.,
voted in the last election but not in the previous three) by simple
deterministic rules is ambiguous and can be suboptimal, as the
procedure lacks the ability to learn structure from a data set.

This has sparked interest in adopting probabilistic approaches in place
of deterministic rules based solely on the voting history [\citet
{Issenberg2012b}]. For instance, \citet{malchow08} promotes a linear
probability model as well as tree-like models such as CHAID [\citet
{chaid}] for political microtargeting. \citet{murray2010} suggest
decision trees as well. \citet{green2012} advocate Bayesian additive
trees [BART, \citet{chipman2010}] and \citet{imai2011} propose to use
classification trees, which they embed in a decision theoretic
framework for optimal planning of GOTV campaigns. Other
state-of-the-art approaches that are used include logistic or probit regression.

%s1.2.1 #&#
\subsubsection{\texorpdfstring{Targeting for turnout.}{Targeting for turnout}}
In the specific case of using probabilistic models for targeting for
turnout, the two tasks of identifying likely voters and likely
supporters from Figure~\ref{figquadrants} coincide. Here, campaigns
are interested in assigning each voter an individual probability to
show up at election day. Based on these estimated probabilities, \citet
{malchow08} reasons that using targeting plans on people with values
around $0.5$ is worthwhile, whereas targeting people with predicted
probabilities near $0$ or $1$ is considered a waste. Given a high
accuracy of the predictions, a person with a predicted probability
close to zero is unlikely to vote, regardless of how compelling the
mobilization message is. A person with a predicted probability of 1 is
going to turn out at the polls anyways, even without the need for extra
persuasion. In both cases, targeting those people would not lead to an
increase in turnout, yet it would consume resources and hence be
wasteful. However, voters with a predicted probability in a ``targeting
range'' around $0.5$ may be ``convincable'' to show up at the polls
using the right incentive. \citet{malchow08} suggests a targeting range
of $[0.3,0.7]$. Clearly, we can be hopeful to sway a person with a
probability of voting of, say, $0.35$, as long as we get the right
message to her. Also, while a person with a probability of, say, $0.68$
might be going to vote without being targeted specifically, it should
not hurt to encourage her a bit more.

%s2 #&#
\section{\texorpdfstring{A new unified statistical framework for voter
targeting.}{A new unified statistical framework for voter
targeting}}
In this paper we introduce a flexible statistical framework for the
task of voter microtargeting and apply it to a (virtual) nonpartisan
GOTV campaign that uses different sets of predictor variables. The main
contribution of this framework is that it allows prediction and
segmentation in a single step. It generalizes two standard models
currently used in political targeting: it encompasses logistic
regression as well as classification trees and also allows for a
combination of both within the same model. We refer to the resulting
framework as {LO}gistic {RE}gression {T}ree (LORET) models. LORET models are very flexible in that, in their
simplest form, they reduce to a majority vote model; they also allow
regression-like modeling with predictors (with small adjustment it
works for all generalized models for binary data such as probit models)
as well as hierarchical partitioning of the feature space under the
same umbrella.

Based on a novel data set of Ohio voters which is prototypical for what
campaigns can buy from data providers, we investigate LORET models of
varying degrees of flexibility and compare them with a particular focus
on the benefits they provide for targeting voters. While we illustrate
LORET models for assessing the probability of turnout only, we are
quick to point out that LORET models can also be used to gauge a
voter's probability of supporting a candidate or cause. We show that
LORET models can have higher predictive accuracy than logistic
regression alone, may lead to better interpretability compared to
classification trees, allow for automatic data-driven creation of voter
profiles, conduct variable selection and allow for inclusion of
substantive knowledge and experience via the logistic model.

This paper is organized as follows. In Section~\ref{secmethods} we
present a statistical framework for voter targeting that combines
logistic regression models with recursive partitioning. Section~\ref{seceval} describes the case study of applying the methods to a
(virtual) nonpartisan GOTV campaign in Ohio that set increasing overall
turnout in the US presidential general election in 2004 as its goal. We
illustrate using the LORET framework in a situation where we have
labeled training data for a sample of eligible voters (Section~\ref{secfcpart}). In Section~\ref{secprofiles} we discuss the creation of
model-based voter profiles for targeting and illustrate how they arise
naturally within the LORET framework. We finish with conclusions and
some general remarks on the usage of LORET in Section~\ref{seccon}.
This paper is accompanied by supplementary material [\citet{supplementa}].

%s3 #&#
\section{\texorpdfstring{LORET: Modeling and predicting voting
behavior.}{LORET: Modeling and predicting voting
behavior}}
\label{secmethods}
Logistic regression and tree-based
methods are popular methods for turnout prediction and voter targeting
[\citet{malchow08}]. Using this as a
backdrop, we introduce a general framework---logistic regression trees
(LORET)---that encompasses and extends these methods. Briefly, the idea
is the
following: Instead of fitting a global logistic regression model to the whole
data, one might fit a collection of local regression models to subsets or
segments of the data (i.e., a segmented logistic regression model) in
order to
obtain a better fit and higher predictive accuracy. Since usually the
``correct'' segmentation is not known, it needs to be learned from the data,
for example, by using recursive partitioning methods.

In what follows we start with the general formulation of logistic regression
models for one or more segments and then show how for more than one
segment the
segmentation can be estimated with recursive partitioning.

%s3.1 #&#
\subsection{\texorpdfstring{Segmented logistic regression.}{Segmented logistic regression}}
Let $y_i \in\{0, 1\}$ denote a Bernoulli random variable for the $i$th
observation, $i=1,
\ldots, N$, and $\mathbf{x}_i$ denote a $(p+1)$-dimensional vector of
$p$ covariates and one intercept,
$(1, x_{i1}, \ldots, x_{ip})^{\top}$. Let us assume there are $r$ (known
or estimated)
disjoint segments in the data. For each segment $k=1,\ldots,r$, we can then
specify a logistic regression model for the relationship between $y$
and $x_1,
\ldots, x_p$ within that segment,
%
%
%e3.1 #&#
%
\begin{equation}
P \bigl(y_{i}=1|x_{i1},\ldots,x_{ip};\bolds{
\beta}^{(k)} \bigr)=\pi_{i}=\frac
{\exp(\x_{i}^{\top}\bolds{\beta}^{(k)})}{1+\exp(\x_{i}^{\top
}\bolds{\beta}^{(k)})},
\label{eqseglogreg}
\end{equation}
where $k = k(i)$ is the segment to which observation $i$ belongs and
$\pi_{i}$ denotes the probability to belong to class ``1'' (e.g.,
``vote${}={}$yes''). The segment-specific parameter vector is $\bolds
{\beta}^{(k)}$ and its estimates are referred to as $\hat{\bolds
{\beta}}^{(k)}$, which can be easily obtained (given the segmentation)
via maximum likelihood [see, e.g., \citet{mccullough89}]. Based on the
associated predicted probabilities, classification can then be done by
%
%
%e3.2 #&#
%
\begin{equation}
\label{eqco2} \hat{y}_{i}(c_{0}) = %
\cases{ 1, &\quad
$\mbox{if } \hat{\pi}_{i}\geq c_{0},$\vspace*{2pt}
\cr
0, & \quad$\mbox{if } \hat{\pi}_{i} < c_{0},$} %
\end{equation}
where $c_{0} \in[0,1]$ is a specific cutoff value (but could, in
principle, also be specified to be different for different segments).

\begin{figure}

\includegraphics{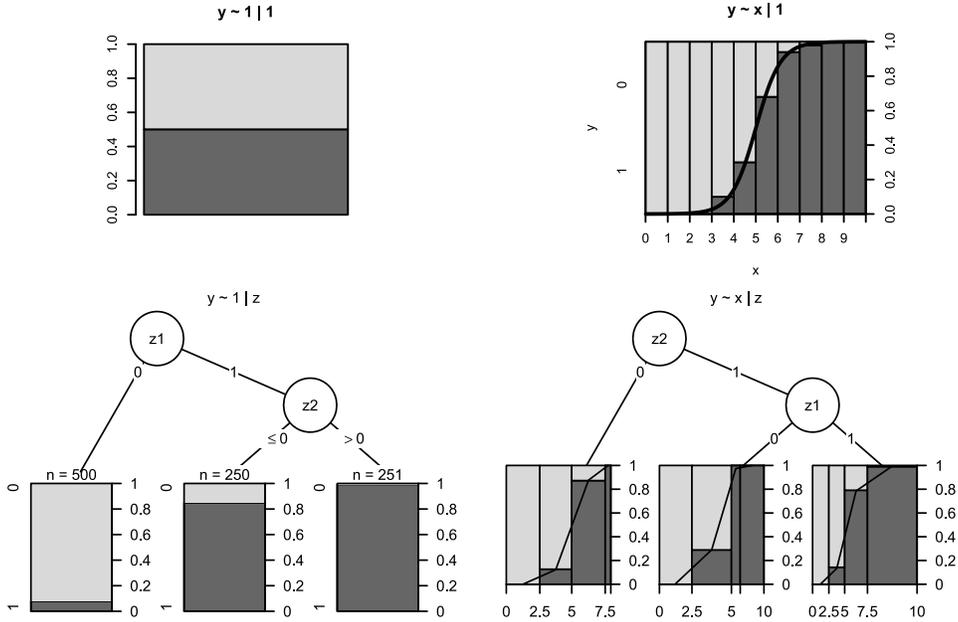}

\caption{A visualization of the different cases of LORET. In the upper
left panel there is the $y\sim1 \Sep1$ LORET, fitting a constant. In
the upper right the $y \sim x \Sep1$ LORET (logistic regression) is
displayed, which is a single function of $x$ for the whole data set.
The lower left panel displays a $y\sim1 \Sep z$ LORET where the data
set is partitioned based on the state of predictor variables $z$ and in
each partition a constant is fitted. In the lower right panel, the
$y\sim x \Sep z$ LORET can be found. Here the data set is again
partitioned based on $z$, but this time a logistic function of $x$ is
fitted in the partitions. Hence, it combines the $y\sim1 \Sep z$ and
$y\sim x \Sep1$ LORET.}\label{figLORETs}
\end{figure}

If there is only a single segment (i.e., a root node and hence a known
segmentation), LORET in (\ref{eqseglogreg}) reduces to a standard
logistic regression model. Here the parameters of the linear
decomposition of the conditional mean of the logit-transformed response
variable $y$ are estimated given the status of $p$ covariates.
Evaluation of the logistic model at the estimated parameter vector
$\hat
{\bolds{\beta}}$ yields the predicted probabilities, $\hat{\pi
}_i$. If the model uses no covariates as regressors, it further reduces
to a majority vote model, that is, a logistic regression model with
only an intercept or simply the relative frequency of class ``1''
transformed to the logit scale. The upper row in Figure~\ref{figLORETs} illustrates majority vote and logistic regression on an
artificial set of data with a single continuous covariate $x$. The
former fits a single constant (the prevalence of ``1''), the latter a
single logistic function of $x$ to the entire data set.

%
%f2 #&#

If there were more than one segment and the segmentation were known,
then LORET can still be simply seen as estimating a maximum likelihood
model from a binomial likelihood in each segment. To estimate it, one
needs to specify a logistic regression model with additional main
effects for the categorical covariates (factors) corresponding to the
segments and the interactions between the segment-factors and the
predictors, but this still falls into the standard theory of
generalized linear models [\citet{mccullough89}].

If the segmentation is unknown, however, it needs to be learned from
the data. Two popular approaches for achieving this are using mixture
models (e.g., mixtures of experts or latent class regression) or
employing some type of algorithmic search method. Recursive
partitioning is a popular example of the latter [with the result often
called a ``tree'', \citet{Zhang2010}]. Trees are usually induced by
splitting the data set along a function of the predictor variables into
a number of partitions or segments. The segments are usually chosen by
minimizing an objective function (e.g., a heterogeneity measure or a
negative log-likelihood) for each segment. The procedure is then
repeated recursively for each resulting partition. This approach
approximates real segments in the data and yields a segmentation for
which maximum likelihood estimation of parameters in each segment can
be carried out, as is done in LORET.

%s3.2 #&#
\subsection{\texorpdfstring{Recursive partitioning.}{Recursive partitioning}}

Let us assume we have an additional, $\ell$-dimensional covariate vector
$\mathbf{z}=(z_1,\ldots, z_\ell)$. Based on these covariates, we
learn the
segmentation, that is, we search for $r$ disjoint cells that partition the
predictor subspace. Depending on whether the logistic model used for
$y$ in each
segment has any covariates or just a constant as regressors, there are two algorithmic
approaches we can use: classification trees and trees with logistic
node models.

%s3.2.1 #&#
\subsubsection{\texorpdfstring{Classification trees.}{Classification trees}}

If the logistic model is an intercept-only model and we have a number of
partitioning variables $z_1, \ldots, z_\ell$, then LORET can be
estimated as a
classification tree. An illustration of a classification tree can be
found in
the lower left panel of Figure~\ref{figLORETs}, where the data is first
partitioned into three subsets and an intercept-only model is fitted to each
subset separately. Hence, in each terminal node the model is a
constant. A wide
variety of algorithms have been developed to fit classification trees, among
them are CHAID [\citet{chaid}], CART [\citet{breiman84}], C4.5 [\citet
{Quinlan1993}], QUEST [\citet{LohShih1997}],
CTree [\citet{hothorn06}] and many others. In this paper, we use CART
and CTree which, respectively, are examples of
tree algorithms that are biased or unbiased in variable selection.

%s3.2.2 #&#
\subsubsection{\texorpdfstring{Trees with logistic node models.}{Trees with logistic node models}}

If there are partitioning variables $\mathbf{z}=(z_1,\ldots, z_\ell)$ as
well as
regressor variables $\mathbf{x} = (1, x_1, \ldots, x_p)$ for the
logistic node model, we get the most general type of LORET, which is a
``model tree.'' The situation is illustrated in the lower right panel
in Figure~\ref{figLORETs}. Like in a classification tree, the data is
first partitioned into subsets. However,
in contrast to a classification tree, separate logistic regressions with
regressors are employed in each terminal node. Thus, the resulting
model tree
essentially combines data-driven partitioning as done by classification
trees with
model-based prediction in a single approach. Different algorithms have been
proposed to estimate model trees with logistic node models, including
the following: SUPPORT
[\citet{Chaudhuri1995}], LOTUS [\citet{ChanLoh2004}], LMT [\citet
{Landwehr2005}] and
MOB [\citet{zeileis08}]. In what follows, we will use the MOB algorithm
with a
logistic node model for estimating the most general version of LORET,
as it
proved to have good properties [\citet{Rusch2012}].

%
%t2 #&#
\begin{table}
\caption{Various instances of LORET}\label{tblmodels}
\begin{tabular*}{\textwidth}{@{\extracolsep{\fill}}lccc@{}}
\hline
\textbf{Method} & \textbf{Regressor variables} &\textbf{Partitioning
variables} &\textbf{Schema}\\
\hline
Majority vote & none & none & $y\sim1 \Sep1$\\
Logistic regression & yes & none & $y\sim x \Sep1$ \\
Classification tree & none & yes &$y\sim1 \Sep z$ \\
Model tree & yes & yes & $y\sim x \Sep z$\\
\hline
\end{tabular*}
\end{table}

To simplify notation and to stress the similarities, we will use a
simple schema to refer to the different LORET types (cf.~Table~\ref{tblmodels} and Figure~\ref{figLORETs}): Majority vote models will be
referred to as $y \sim1 \Sep1$, global logistic regression models as
$y \sim x \Sep1$, classification tree models as $y \sim1 \Sep z$ and
full LORET model as $y \sim x \Sep z$.

The LORET framework can be employed for various tasks during a voter
targeting or get-out-the-vote campaign. To illustrate the usage of
LORET in a campaign's voter targeting strategy, we use a unique,
proprietary data set from the 2004 general presidential election in
Ohio, USA.

%s4 #&#
\section{\texorpdfstring{Case study: Get-out-the-vote in Ohio.}{Case study: Get-out-the-vote in Ohio}}
\label{seceval}
We apply our methodology to a (fictional) nonpartisan get-out-the vote
campaign in Ohio, USA, whose goal it is to increase voter turnout. We
choose Ohio because it has proven to be a pivotal state in about every
US presidential election since 1964. Also, in every US presidential
election since 2000, the difference between the Republican and
Democratic candidates has been equal or less than 4\%, making it a top
battleground state in every recent election. The campaign we describe
pertains to the 2004 US presidential election.

Our data set originates from a data vendor who adds value to public
records by collecting, maintaining, updating and expanding upon public
data. In the US, vendor voter data typically includes the name,
address, phone, gender, party affiliation, age, vote history (elections
that each voter voted) or ethnicity. US data vendors standardize the
data by each state or county and by adding other potentially relevant
behavioral information such as income, type of occupation, education,
presence of children, property status (rental or owning) and charities
that the person donated to.

%s4.1 #&#
\subsection{\texorpdfstring{Data description.}{Data description}}
For illustration we use a proprietary data set\footnote{We are not at
liberty to share the whole data set but included a snapshot of 6544
anonymized records to make our results comprehensible and for further
research, see the supplementary material [\citet{supplementa}].} which was provided by one of the leading
nonpartisan data vendors in the industry. The data set consists of
records from $19\mbox{,}634$ eligible and registered voters from Ohio. It
includes a total of $77$ variables, many of which are sociodemographic
categorical variables like gender, job category or education level. The
data set also contains records on past voting behavior from 1990 to
2004 in general elections, primary or presidential primary elections
and other elections, all coded as binary variables---that is, voted
(``yes'') or not (``no''). We added three composite or aggregate
variables: the raw count of elections a person attended, the number of
elections a person attended since registering and the relative
frequency of attended elections since registering. After removal of
missing values and inconsistent entries (366 cases) there are a total
of $N=19\mbox{,}634$ records with $80$ variables per record. The
variable we
want to predict is the individual turnout likelihood in the 2004 US
presidential election.

%s4.2 #&#
\subsection{\texorpdfstring{Two sets of predictors: Voting history only vs. kitchen
sink data.}{Two sets of predictors: Voting history only vs. kitchen
sink data}}
The data available to campaigns can vary vastly. Some campaigns have a
huge number of variables on millions of eligible voters available, as
was the case with President Obama's re-election campaign in 2012
[Project ``Narwhal'', \citet{Issenberg2012}].
Smaller campaigns may have more limited information available. For all
cases, however, the literature on voter targeting suggests that the
most commonly used piece of information is the person's voting history
[\citet{malchow08}], although often taking into account a person's age
[\citet{malchow08,karp2008}] is recommended. One of the goals of this
case study is to investigate whether including additional information
(besides a person's voting history and age) into the targeting model is
beneficial. To that end, we compare and contrast two sets of predictors:
\begin{itemize}
\item The first set employs the standard information used by many
campaigns, which is also recommended in the literature. These standard
variables are a person's voting history, recorded over the last four
elections, and age. We call this set ``$s$'' for ``standard.''
\item The second set contains all other variables available, that is,
the ``kitchen sink.'' In our case this includes variables like gender,
occupation, living situation, party affiliation, party makeup of the
household (``partyMix''), position within the family (``hhRank'' and
``hhHead''), donations for various causes, education level, relative
frequency of attended elections so far (``attendance'') and many
others. These variables constitute a set of additional variables,
labeled ``$e$'' for ``extended.''
\end{itemize}

%s4.3 #&#
\subsection{\texorpdfstring{Model specification for the Ohio voters.}{Model specification for the Ohio voters}}
The combination of the two variable sets with the different LORET
models leads to model specifications as displayed in Table~\ref{tblmodelsBig}. The models
either employ only the standard set of variables or the combination of
the standard and the extended set. For unpartitioned models, the
parameters are
estimated with maximum likelihood. If a partition is induced, we learn
it with three different algorithms (CART, CTree and MOB), depending on
the nature of
the node model. Please note that if age is specified as a parameter in
the logistic model part (i.e., for models $y\sim s \Sep1$, $y\sim s+e
\Sep1$ and $y\sim s \Sep e$), a quadratic effect will be used [based
on goodness-of-fit considerations; see also \citet{parry2008}].

%
%t3 #&#
\begin{table}
\caption{LORET versions combined with the two variable groups and the
algorithms used to estimate the partition. The standard variable set of
age and voting
history is labeled ``$s$'' and the set of additional variables with
``$e$'' (hence all variables together are ``$s+e$'')}\label{tblmodelsBig}
\begin{tabular*}{\textwidth}{@{\extracolsep{\fill}}lccc@{}}
\hline
\textbf{LORET} & \textbf{Regressor variables} &\textbf{Partitioning
variables} & \textbf{Partitioning
algorithm} \\
\hline
$y \sim1 \Sep1$ & none & none & -- \\
$y \sim s \Sep1 $ & $s$ & none & --\\
$y \sim s+e \Sep1$ & $s + e$ & none & --\\
$y \sim1 \Sep s$ & none & $s$ & CART, CTree\\
$y \sim1 \Sep s+e$ & none & $s + e$ & CART, CTree\\
$y \sim s \Sep e$ & $s$ & $e$ & MOB\\
\hline
\end{tabular*}
\end{table}

All recursive partitioning algorithms that we employ allow for tuning
with metaparameters. These tuning parameters can be used to avoid
overfitting of the tree algorithms and control how branchy the tree
becomes. Quite generally, it can be said that the less branchy a tree
is, the less prone it is to overfitting. In the algorithms we can use a
higher number of observations per node, a lower tree depth and a
stricter split variable selection criterion that all lead to smaller
trees. At the same time the specification of metaparameters should
grant enough flexibility for the algorithm to approximate a complex
nonlinear relationship in the data.

For CART the maximal depth of the tree and the minimum number of
observation per node (minsplit) are available to control the tree
appearance. We use a maximal tree depth of $7$ and a minsplit of $100$
(which corresponds to roughly $0.5\%$ of the observations). For CTree
and MOB the significance level of the association or stability tests,
respectively, and the minimum number of observation per node can be
used to tune the algorithm and pre-prune the trees. We employ a global
significance level of $\alpha=1\times10^{-6}$. This is sensible since
the high number of observations might easily lead to significant
results mainly due to the sample size. Hence, we reduce the chance of
``false positive'' selection of a split variable or split point by
specifying a low significance level. This also functions as ``automatic
regularization,'' as the test statistics used to decide whether to
split a node have to become larger the larger the tree becomes. For
minsplit we use $100$ for CTree (the same as for CART) and $1000$ for
MOB which enables reliable estimation of the node model. Please note
that the results were not sensitive to the choice of metaparameters.
For CART, we explored depths from $3$ to $20$. For the global
significance levels of CTree and MOB, we explored values of $0.0001$,
$0.0005$, $0.001$, $0.005$, $0.01$, $0.05$ and $0.1$. For the minimum
number of observations a node must contain we explored values of $20$,
$50$, $100$, $150$, $200$, $250$ and $500$ for all methods. For these
choices of depth, number of observations per node and significance
level, the results were very similar.

In what follows we illustrate targeting based on LORET.
We start with voter targeting in a setting where proxy data about the
voting behavior for a sample of individuals in the upcoming election is
available (e.g., from a poll). We then highlight the use of LORET for
the creation of voter profiles.

%s4.4 #&#
\subsection{\texorpdfstring{Predicting individual turnout.}{Predicting individual turnout}}
\label{secfcpart}

Typically the individual turnout is only known after the election is
over. This makes the application of supervised procedures like LORET
during or before the election challenging, since supervised procedures
rely on a labeled training set in order to derive predictions. It is
therefore imperative for campaigns to obtain labeled proxy data prior
to closing of the election booths that most accurately resembles the
true outcome. These data will often arrive from carefully designed,
reliable, repeated polls about voting intention.
Information gathered this way can be turned into labels for training a
supervised classification model. For our virtual campaign to mobilize
Ohio voters, we simulate this by estimating LORET via a training set
drawn randomly (see also further below) from the entire data. Other
proxies that can be used are past election results. (We also
investigated our method with using the previous presidential election
as proxy variable. The predictive accuracy was low---around 0.72, with
majority vote having an accuracy of 0.7. We concluded that this is no
viable alternative to surveys of people's voting intentions, so we
refrained from presenting the results in the main paper. The supplementary material [\citet{supplementb}]
contains a thorough account of that analysis.)

%s4.4.1 #&#
\subsubsection{\texorpdfstring{Learning and test samples via
bootstrapping.}{Learning and test samples via
bootstrapping}}
We simulate the targeting situation based on labeled training data by
drawing a bootstrap sample [see, e.g., \citet{hastie06}], that
is, a learning set of size $N$ which is sampled randomly (with
replacement) from the entire set of data and use this as our training
set. To the learning set we fit a LORET model and use the model to
predict the out-of-bag (oob) test set which consists of observations
that were not part of the learning sample and thus basically treating
them as having an unknown label. To evaluate and compare the different
models, we employ the benchmarking framework of \citet{hothorn2005}.
Ten folds of learning and test samples $f= 1,\ldots, 10$ are used. To
provide a further benchmark, we also train and evaluate all models on
the whole data set. This allows us to gauge the tendency of a model to
overfit as well as how close out-of-bag and in-sample performance
are.\looseness=-1

%s4.4.2 #&#
\subsubsection{\texorpdfstring{Measuring predictive accuracy.}{Measuring predictive accuracy}}
For each method, we assess the classification accuracy ($\mathit{acc}_f$) on
each oob test set $f$ at a given cutoff value $c_0=0.5$ (for
simplicity, we use the same cutoff value of $0.5$ for all segments $k$).
To estimate overall predictive accuracy, we use the average over all
bootstrap samples $\overline{\mathit{acc}}$. %=\frac{1}{F}\sum_{f=1}^{F}
When using the full data set as training and test set (i.e., in-sample
performance), we denote the accuracy by $\mathit{acc}_0$.

Furthermore, we use the ROC curve for model comparison. It displays the
false positive rate vs.~the true positive rate. For a given threshold
value, we average the ROC curves across all bootstrap samples. The area
under the ROC curve\vadjust{\goodbreak} for oob set $f$, $\mathit{auc}_f$, serves as a
cutoff-independent measure of classification accuracy and we calculate
it via the Wilcoxon statistic [\citet{Wilcoxon1945}]. Once again, we
average it over all bootstrap samples ($\overline{\mathit{auc}}$) %=\frac{1}{F}
and use $\mathit{auc}_0$ to denote the in-sample area under the curve. For all
the classification measures above, higher values imply better
predictive capability.
By using simultaneous pairwise confidence intervals [using Tukey's
all-pairwise comparison contrasts and controlling for the family-wise
error rate, cf. \citet{multcomp}] around the differences in predictive
accuracy and AUC between two models, we assess whether the real
differences can be judged to be different from zero (95\% confidence).
To account for the dependency structure of bootstrap samples, we center
the accuracies beforehand [see \citet{hothorn2005}].

%
%f3 #&#
\begin{figure}[b]

\includegraphics{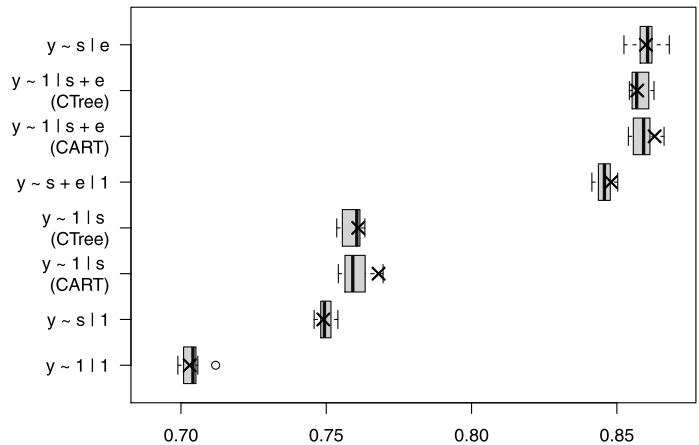}

\caption{Accuracies for LORET models with different sets of predictors:
Accuracy boxplots at a cutoff of 0.5 for all 10 out-of-bag samples for
each LORET instance are displayed. The cross denotes the in-sample
prediction accuracy of each of these models ($\mathit{acc}_0$).%The upward
%triangle denotes the forecast accuracy of the $y00 \sim s00$ models
%and the downward triangle for the $y00 \sim s04$ models.
}\label{figboxplots}
\end{figure}

%s4.4.3 #&#
\subsubsection{\texorpdfstring{Results.}{Results}}
\label{secres}

\begin{figure}

\includegraphics{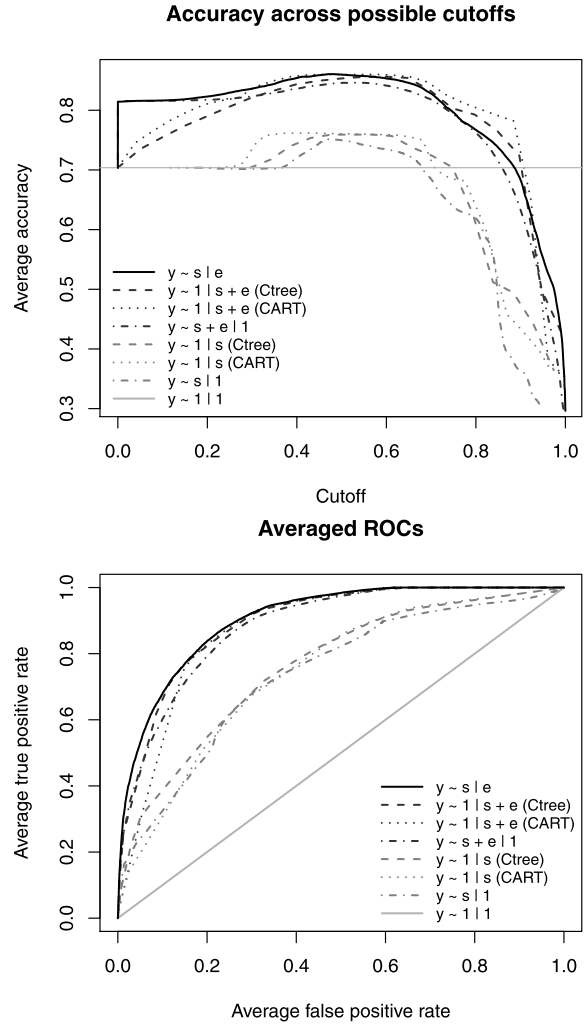}

\caption{Performance indicators for different models. The upper panel
features the average accuracies for the range of different cutoffs for
the various LORET instances (for majority vote the average accuracy is
displayed as a constant). The lower panel features the averaged
receiver operating characteristic (ROC) curve for the different models.
Threshold averaging has been used for all methods except majority
vote.}\label{figROCs}
\end{figure}

Looking at the upper part of Figure~\ref{figboxplots}, which shows
boxplots of the predictive accuracy for the bootstrap samples as well
as the in-sample accuracy (denoted by a cross) at a cutoff value of
$0.5$, one can see quite clearly how the different models from
Table~\ref{tblmodelsBig} behave for our data. First, using both
variable sets (the standard set and the extended set together) leads to
a large improvement in predictive accuracy as compared to just using
the standard set. Interestingly, the improvement of using both the
``$s$'' and ``$e$'' variables over using only ``$s$'' is bigger than
the improvement of using only ``$s$'' over using no covariates at all
(cf.~Figure~\ref{figboxplots}). Second, LORET versions that employ
recursive partitioning perform better than global regression models
alone. This holds for using only the standard variable set as well as
the combination of the extended and standard sets. This can also be
seen in Figure~\ref{figROCs} which displays the average classification
accuracies as\vadjust{\goodbreak} a function of different cutoff values in the upper panel
and the mean ROC curves in the lower panel (averaged over the $F=10$
out-of-bag samples).
%

%
%f4 #&#

Table~\ref{tblsumm} gives a detailed summary of the different
performance measures for all models. The benchmark of the naive model
$y\sim1 \Sep1$ is an average prediction accuracy of $\overline
{\mathit{acc}}=70.36\%$ and an AUC of $\overline{\mathit{auc}}=0.5$, averaged over all
test sets.

Global logistic regression models $y\sim s \Sep1$ and $y\sim s+e \Sep
1$ display improved performance ($\overline{\mathit{acc}}=74.97\%$ and
$\overline
{\mathit{auc}}=0.740$ for the standard set and $\overline{\mathit{acc}}=84.57\%$ and
$\overline{\mathit{auc}}=0.886$ for the combined set) with a huge improvement of
the model that uses both variable sets.

Both classification tree algorithms, CART and CTree, used to estimate
$y\sim1 \Sep s$ and $y\sim1 \Sep s+e$ result in a generally better
performance compared to logistic regressions, both on the standard set
of predictors as well as for combining the standard and the extended
set. Their performance peaks for the combined set with values of
$\overline{\mathit{acc}}=85.96\%$ and $\overline{\mathit{auc}}=0.878$ for $y\sim1 \Sep
s+e$ (CART) and $\overline{\mathit{acc}}=85.78\%$ and $\overline{\mathit{auc}}=0.898$ for
$y\sim1 \Sep s+e$ (CTree).

For the LORET that uses the standard set of predictors as the model in
the terminal nodes of the tree and the extended set of predictors for
partitioning, that is, $y\sim s \Sep e$ result values of $\overline
{\mathit{acc}}=85.98\%$ and $\overline{\mathit{auc}}=0.906$, respectively. %
%and, at this cutoff, the highest mean accuracy.} %(although the latter
%is not statistically different from the other tree models).

%
%t4 #&#
\begin{table}
\tabcolsep=0pt
\caption{Summary of performance indicators for each LORET instance. For
the bootstrap samples,
$\overline{\mathit{auc}}$ means the area under the ROC curve averaged over all
$10$ out-of-bag test sets.
$\overline{\mathit{acc}}$ is the overall classification accuracy averaged over
all test sets and $\operatorname{se}(\mathit{acc})$ its standard error.
Complexity is given as the number of estimated parameters per segment
(terminal node) $p+1$ and the
median number of segments $\tilde{r}$. For the full sample models
(fitted and evaluated on all observations),
the accuracy is given by $\mathit{acc}_0$, the AUC by $\mathit{auc}_0$ and the number of
terminal nodes and cofficients
in each node by $r_0$ and $p_0+1$, respectively}\label{tblsumm}
\begin{tabular*}{\textwidth}{@{\extracolsep{\fill
}}lcccd{2.0}d{2.1}ccd{2.0}d{2.0}@{}}
\hline
&\multicolumn{5}{c}{\textbf{Bootstrap samples}} & \multicolumn
{4}{c@{}}{\textbf{Full sample}}\\ [-6pt]
&\multicolumn{5}{c}{\hrulefill} & \multicolumn{4}{c@{}}{\hrulefill
}\\
Method&\multicolumn{1}{c}{$\bolds{\overline{\mathit{acc}}}$}&\multicolumn
{1}{c}{$\bolds{\operatorname{se}(\mathit{acc})}$}&\multicolumn{1}{c}{$\bolds{\overline{\mathit{auc}}}$}&
\multicolumn{1}{c}{$\bolds{p+1}$}&\multicolumn{1}{c}{$\bolds{\tilde
{r}}$}&\multicolumn{1}{c}{$\bolds{\mathit{acc}_0}$}&\multicolumn
{1}{c}{$\bolds{\mathit{auc}_0}$}&\multicolumn{1}{c}{$\bolds
{p_0+1}$}&\multicolumn{1}{c@{}}{$\bolds{r_0}$}\\
\hline
$y\sim1 \Sep1$ & 0.704 & 0.004 & 0.500 & 1 & 1.0 & 0.703 & 0.500 & 1
& 1 \\[3pt]
$y\sim s \Sep1$ & 0.750 & 0.002 & 0.740 & 8 & 1.0 & 0.749 & 0.739 & 8
& 1 \\
$y\sim1 \Sep s$ (CTree) & 0.759 & 0.004 & 0.765 & 1 & 15.0 & 0.761 &
0.762 & 1 & 14 \\
$y\sim1 \Sep s$ (CART) & 0.760 & 0.005 & 0.745 & 1 & 28.5 & 0.768 &
0.746 & 1 & 27 \\[3pt]
$y\sim s+e \Sep1$ & 0.846 & 0.003 & 0.886 & 57 & 1.0 & 0.848 & 0.888
& 57 & 1 \\
$y\sim1 \Sep s+e$ (CTree) & 0.858 & 0.003 & 0.898 & 1 & 18.0 & 0.857
& 0.898 & 1 & 18 \\
$y\sim1 \Sep s+e$ (CART) & 0.860 & 0.004 & 0.878 & 1 & 23.5 & 0.863 &
0.886 & 1 & 23 \\
$y\sim s \Sep e$ & 0.860 & 0.004 & 0.906 & 8 & 9.5 & 0.860 & 0.909 & 8
& 8 \\
\hline
\end{tabular*}
\end{table}
%

%f5 #&#
\begin{figure}

\includegraphics{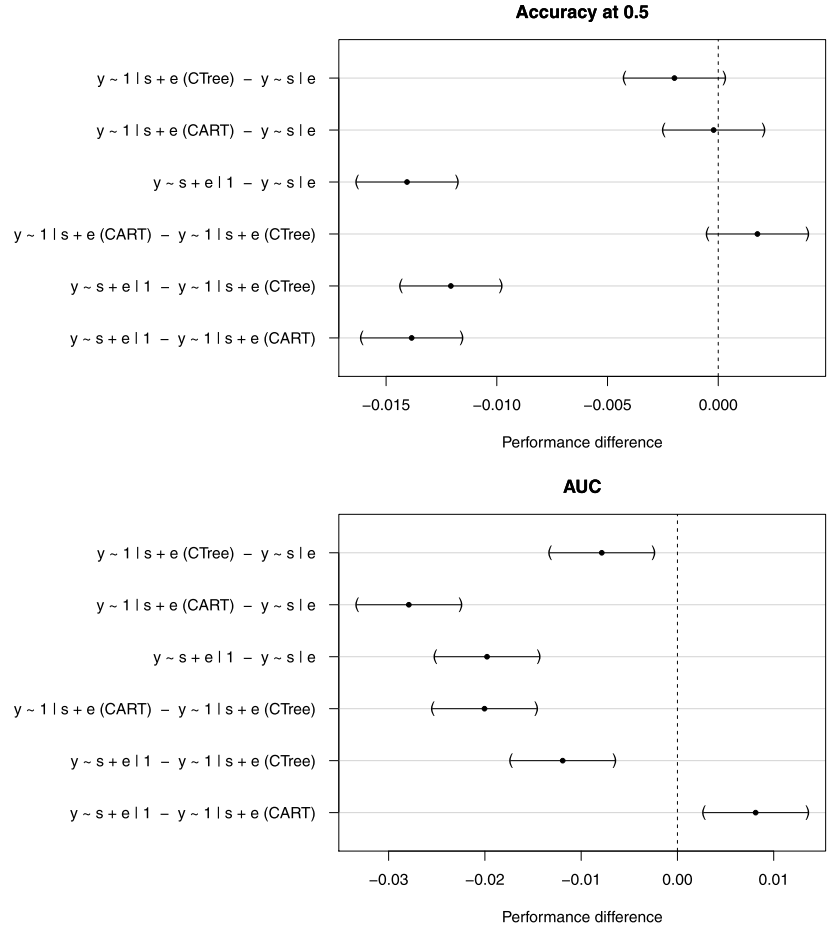}

\caption{Simultaneous pairwise confidence intervals of the differences
of mean accuracies at a cutoff 0.5 over the 10 out-of-bag samples
(upper panel) and differences of the average area under the ROC curve
(AUC) over the 10 out-of-bag samples (lower panel) for all methods
employing the combination of the standard and extended variable
set.}\label{figcis}
\end{figure}

The performance differences of models using only standard variables and
models employing both the standard and the extended variable sets are
evident (see Table~\ref{tblsumm} and Figure~\ref{figboxplots}).
Making use of the additional variables leads to highly improved performance.

However, the differences among the models employing the combined set
themselves (especially between the global logistic regression model and
partitioned models) are not that strong. Therefore, to establish a
region of performance differences that could be expected if all models
performed equally well, we calculated simultaneous $95\%$-confidence
intervals of all pairwise performance differences between the models
that use the combined set of variables based on their accuracy as well
as AUC. The former can be found in the upper panel of Figure~\ref{figcis},
the latter in the lower panel. We can see that the global
logistic regression model performs significantly worse than the
partitioned models ($\alpha=0.05$). The tree methods perform best in
terms of accuracy and their intervals overlap. In contrast, in terms of
the cutoff free measure AUC, the $y\sim s \Sep e$ LORET significantly
outperforms all other methods.
\newpage

%s4.5 #&#
\subsection{\texorpdfstring{Voter segmentation (``Voter profiles'').}{Voter segmentation (``Voter profiles'')}}
\label{secprofiles}

``Voter profiles'' are descriptions of a voter or set of voters that
may include demographic, geographic and psychographic characteristics,
as well as voting patterns and voting history.
Voter profiles are popular in targeting efforts by campaigns, as they
allow to break the complexity of all the available data down into a
small number of key characteristics that can easily be acted upon. Key
demographic variables are gender, income, age and education. A famous
example of a voter profile is the ``soccer mom'' [\citet{Susan1999}].

Multivariate voter profiles arise naturally from the LORET framework
and the resulting profiles have two distinct benefits:
On the one hand, the voter profiles are automatically created by a
data-driven procedure, as tree-based methods algorithmically segment
the data into mutually exclusive subsets. The segmentation is based on
predictor variables in a well-defined fashion and the selection of
important predictors is (usually) done automatically.
On the other hand, logistic regression and trees with logistic node
models are able to express an individual probability for each voter to
turn up at the polls by including regressor variables in the logistic
model and thus further differentiate the predicted probability between
people in a segment. This way logistic regressions and model trees can provide
individual predictions rather than a single prediction for a given
profile. Furthermore, the estimates of the logistic model and/or the
decision rules of the trees offer additional insight into the dynamics
of voting behavior.

As case in point, consider the most general LORET, $y \sim s \Sep e$.
We have shown in the previous section that it has high accuracy and AUC
for this data set.
To derive voter profiles based on this model, we fit the logistic
regression tree to the whole data set. The decision rules for building
the segments and the coefficients for the logistic regression model in
each terminal node can be found in Table~\ref{tbltreeres}.

We can see that the segmentation is driven by only four variables, the
party composition of the household for each voter (``partyMix''), the
relative frequency of attended elections (``attendance''), the rank of
the individual in the household (``hhRank,'' with ``1'' being highest
and ``3$+$'' being lowest) and whether the person is the head (``H'') or
a member (``M'') of the household (``hhHead''). Hence, most
partitioning variables are concerned with the household structure
rather than with individual-level variables. This underlines a streak
of literature that emphasizes the importance of the household for
voting behavior [e.g., \citet{cutts2009}].
\begin{sidewaystable}
\tabcolsep=0pt
\tablewidth=\textwidth
\caption{A tabular representation of the terminal nodes for the $y\sim
s \Sep e$ LORET for the whole Ohio voter data set.
The first column lists the terminal node numbers. The next four
columns list the partitioning variables (party mix, attendance,
household rank and household head) and the split point
(if any). The last eight columns list the coefficients (upper row) and
standard errors (lower row) for the fitted
logistic models in the nodes. Please note that the values for the
quadratic effect of age have been multiplied by
100 for readability.}
\label{tbltreeres}
\begin{tabular*}{\textwidth}{@{\extracolsep{\fill}}lcccccccccccc@{}}
\hline
&\multicolumn{4}{c}{\textbf{Partitioning variables}}&\multicolumn
{8}{c}{\textbf{Regressor variables}}\\[-6pt]
&\multicolumn{4}{c}{\hrulefill}&\multicolumn
{8}{c@{}}{\hrulefill}\\
\multicolumn{1}{@{}l}{\textbf{Segment}} & \multicolumn{1}{c}{\textbf
{partyMix}} &\multicolumn{1}{c}{\textbf{attend.}} &
\multicolumn{1}{c}{\textbf{hhRank}}&\multicolumn{1}{c}{\textbf
{hhHead}}&\multicolumn{1}{c}{\textbf{const.}}& \multicolumn
{1}{c}{\textbf{gen00}} &
\multicolumn{1}{c}{\textbf{gen01}} &\multicolumn{1}{c}{\textbf{gen02}}&
\multicolumn{1}{c}{\textbf{gen03}}& \multicolumn{1}{c}{\textbf
{ppp04}}& \multicolumn{1}{c}{\textbf{age}} & \multicolumn
{1}{c@{}}{$\bolds{\mathrm{age}^2}\bolds{\cdot 100}$} \\
\hline
\phantom{0}2 & unknown & -- & -- & -- & $-\infty$ & $0.000$ & $0.000$
& $0.000$ &
$0.000$ & $0.000$ & $0.000$ & $0.000$ \\
& & & & & (--.--) & (--.--) & (--.--) & (--.--) & (--.--) & (--.--) &
(--.--) & (--.--) \\
\phantom{0}6 & allD & $\leq$0.48 & -- & -- & $ 0.508$ & $ 0.840$ &
$-1.474$ & $
0.287$ & $-0.750$ & $ 0.442$ & $ 0.054$ & $-0.038$ \\
& & & & & $(0.623)$ & $(0.269)$ & $(0.212)$ & $(0.212)$ & $(0.212)$ &
$(0.231)$ & $(0.024)$ & $(0.022)$ \\
\phantom{0}7 & allR, onlyRorD & $\leq$0.48 & -- & -- & $ 0.427$ & $
0.740$ &
$-0.465$ & $ 0.756$ & $-0.075$ & $ 0.708$ & $ 0.011$ & $-0.004$ \\
& & & & & $(0.660)$ & $(0.239)$ & $(0.174)$ & $(0.185)$ & $(0.177)$ &
$(0.169)$ & $(0.028)$ & $(0.027)$ \\
\phantom{0}8 & allR, allD, & $>$0.48 & -- & -- & $ 2.760$ & $ 0.277$ & $-1.164$
& $ 0.352$ & $-1.890$ & $-0.952$ & $ 0.035$ & $-0.017$ \\
& onlyRorD & & & & $(0.948)$ & $(0.339)$ & $(0.352)$ & $(0.379)$ &
$(0.604)$ & $(0.354)$ & $(0.025)$ & $(0.021)$ \\
10 & noneRorD, noneD, & -- & -- & 3$+$ & $ 4.057$ & $ 0.781 $ & $ 0.591
$ & $ 1.249$ & $ 1.520$ & $ 0.677 $ & $-0.250$ & $ 0.272$ \\
& noneR, legal & & & & $(0.797)$ & $(0.128)$ & $(0.203)$ & $(0.165)$ &
$(0.214)$ & $(0.212)$ & $(0.052)$ & $(0.076)$ \\
12 & noneRorD, noneD, & -- & $<\!3+$ & H & $-3.630$ & $ 1.415$ & $-0.010$
& $ 1.521 $ & $ 2.218$ & $ 1.694$ & $ 0.116$ & $-0.108$ \\
& noneR, legal & & & & $(0.339)$ & $(0.079)$ & $(0.111)$ & $(0.105)$ &
$(0.167)$ & $(0.223)$ & $(0.013)$ & $(0.012)$ \\
13 & noneRorD, noneD, & -- & $<\!3+$ & M & $-1.868$ & $ 1.217$ & $ 0.086$
& $ 1.081$ & $ 1.700$ & $ 1.603$ & $ 0.079$ & $-0.078$ \\
& noneR, legal & & & & $(0.428)$ & $(0.113)$ & $(0.148)$ & $(0.133)$ &
$(0.193)$ & $(0.262)$ & $(0.019)$ & $(0.021)$ \\
\hline
\end{tabular*}
\end{sidewaystable}
Note that none of the commonly used demographic variables like gender,
education or income plays a role in our tree. We therefore have voter
profiles that suggest to look at whether a person comes from a
household where all members are Democrats, all members are Republican
or Democrats or a combination of both, or unknown composition, and all
with potentially unaffiliated voters in the household. Additionally,
our model suggests that one needs to consider the rank of each person
in the household and how often the person went voting in the past. The
segmentation then gives rise to different logistic models that provide
additional targeting suggestions for a campaign based on the
coefficients (cf. Table~\ref{tbltreeres} and the predicted
probabilities of each individual person in Table~\ref{tblrankedList}).

%
%t5 #&#

\begin{figure}

\includegraphics{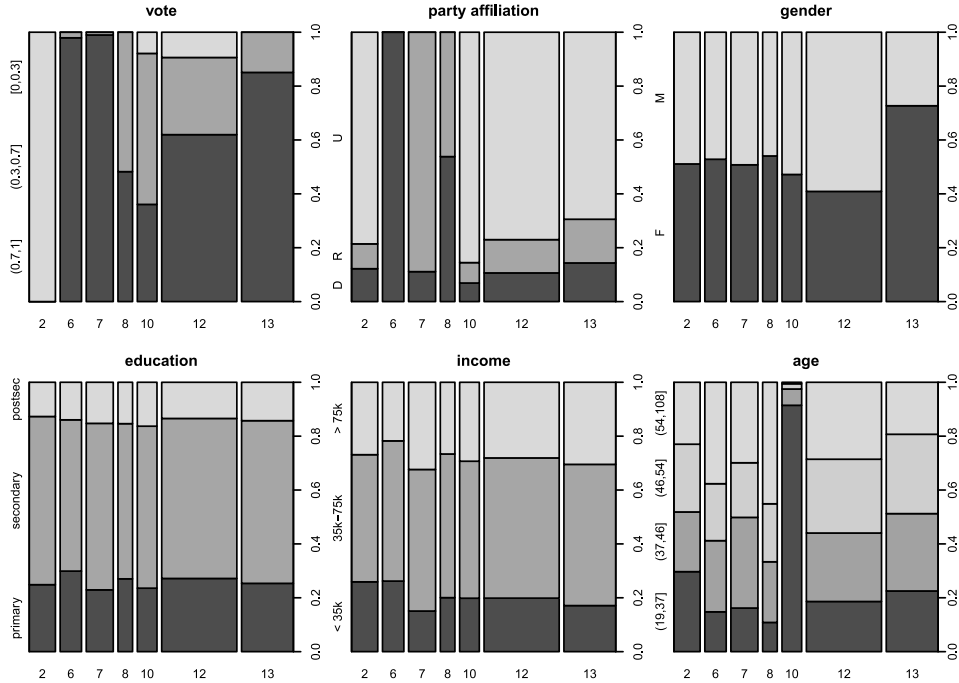}

\caption{Spineplots of the marginal distribution of important voter
profile variables for all segments (the segment number is on the
$x$-axis). The variables are vote (a categorization of the predicted
voting probability: ``likely'' $(0.7,1]$, ``undecided'' $(0.3,0.7]$,
``unlikely'' $[0,0.3]$), party affiliation (``D'' for Democrat, ``R''
for Republican and ``U'' for unaffiliated/independent), gender (``M''
for male, ``F'' for female), education level (``primary,''
``secondary,'' ``postsec''), yearly income (``$<$35k,''
``35k--75k'' and ``$>$75k'') and age category.} \label{figprofiles}
\end{figure}

The results of the segmentation can be used to build more refined voter
profiles by looking at the marginal distribution of different variables
as displayed in Figure~\ref{figprofiles}. These profiles also allow to
derive strategic implications for a targeting campaign. For instance,
for all individuals for whom ``partyMix'' is unknown (segment 2), we
find the predicted probability to vote is near zero [actually a case
where for a linear combination of predictors we have only one level of
the outcome, or quasi-complete separation, \citet{albert84}]. We
further see that people in this segment are mostly independent voters
(78.6\%), relatively often between 19 and 36 year-old individuals (29\%), have a secondary education (62.4\%) and earn between 35,000 and
75,000 USD a year (47.2\%).

%
%t6 #&#
\begin{table}
\tabcolsep=4pt
\caption{An example for a targeting list based on predicted
probabilities of 10 randomly selected individuals for the $y \sim s+e$
LORET of the Ohio voter data file}\label{tblrankedList}
\begin{tabular*}{\textwidth}{@{\extracolsep{\fill}}ld{2.0}cccd{2.2}ccc@{}}
\hline
& & & & & \multicolumn{1}{c}{\textbf{Abs. freq.}} &\multicolumn
{1}{c}{\textbf{Rel. freq.}}
& & \\
\multicolumn{1}{@{}l}{$\bolds{\hat{\pi}_i}$} & \multicolumn
{1}{c}{\textbf{Segment}} & \multicolumn{1}{c}{\textbf{Age}} &
\multicolumn{1}{c}{\textbf{Income}} & \multicolumn{1}{c}{\textbf
{Education}} & \multicolumn{1}{c}{\textbf{votes}} &\multicolumn
{1}{c}{\textbf{votes}} &
\multicolumn{1}{c}{\textbf{Gender}} & \multicolumn{1}{c@{}}{\textbf
{Party}} \\
\hline
1.00 & 13 & 60.02 & D & C & 6.00 & 0.14 & F & R \\
0.95 & 12 & 44.54 & E & D & 4.00 & 0.15 & M & U \\
0.93 & 7 & 63.42 & D & B & 21.00 & 0.48 & F & R \\
0.92 & 13 & 51.30 & I & E & 14.00 & 0.32 & F & U \\
0.88 & 8 & 22.97 & C & D & 7.00 & 0.50 & F & D \\
0.52 & 14 & 27.03 & C & B & 3.00 & 0.12 & F & U \\
0.44 & 14 & 30.24 & E & B & 1.00 & 0.07 & F & U \\
0.41 & 13 & 25.64 & F & C & 3.00 & 0.00 & F & U \\
0.18 & 12 & 23.69 & D & B & 0.00 & 0.00 & F & U \\
0.00 & 2 & 47.39 & F & C & 1.00 & 0.12 & F & U \\
\hline
\end{tabular*}
\end{table}

The most likely voters can be found in segment~7 (mean and median
predicted voting probability of $0.908$ and $0.925$, resp.) and
13 (mean and median predicted probability of $0.861$ and $0.939$,
resp.). Segment 7 has the highest percentage of likely voters
(99\%, see Figure~\ref{figprofiles}) and consists of people who come
from households that either are comprised only of Republicans or of
both Republicans and Democrats and who went voting less than 48\% of
the times. The people in this segment are most often between 36 and 46
years of age (33.7\%) or older than 55 (29.4\%), declared Republican
voters (88.9\%) and often head of a household (56.6\%). With 32.2\%,
segment 7 has the highest proportion of people with high income (more
than 75,000 USD a year) compared to all other segments.
In Segment 13 are people from households with at least one, but
predominantely only unaffiliated voters in the household and whose
household rank is 2. Roughly three quarters (72\%) in this segment are
women. Together with the household rank of 2 this points toward this
being a segment of spouses or partners (typically wives). The most
frequent age group in this segment is 46--54 (31.1\%).
Age has an interesting differential effect in these two segments of
likely voters: When looking at the coefficients of the logistic
regression model---recall that we specified a quadratic effect---we see
that for segment 7 the turning point is at a high age of 70, but for
segment 13 it already appears at 51.1 years.

%
%f6 #&#z

With respect to the mobilization of voters who are undecided as to
whether they will turnout, segments 10 and 8 are most interesting.
As the top left panel in Figure~\ref{figprofiles} shows, segment 10 is
the segment with the highest proportion of ``undecided'' voters (56.04\%).
These voters are from a household with at least one independent or
unaffiliated member and have a household rank of 3 or more. This
segment is special insofar as it contains nearly exclusively young
people (between 19 and 26, 91.2\%) that describe themselves as
unaffiliated voters (85.5\%). This segment consists of the highest
proportion of people with post-secondary education (16.3\%). It
collects young, unaffiliated voters whose predicted probabilities fall
into the targeting range in more than 50\% of the cases. In contrast,
the second ``undecided'' segment, segment 8 (51.7\% undecided), is
characterized by people who are supporters of either the Republican or
the Democratic Party in near equal numbers. Additionally, this segment
has the highest proportion of elderly voters (44.7\% are over 55). This
segment would best be described as elderly, partisan voters who tend to
be predicted as being undecided.

As an alternative to the aggregated view with voter profiles, a
campaign can also use the nonaggregated predicted probabilities by
generating a turnout or support probability for each voter in the
database. Table~\ref{tblrankedList} shows an example with 10 randomly
chosen individuals. We list their predicted probabilities together with
the according realizations of some additional variables. With such a
list it is up to the individual campaign to decide how they eventually
want to rank the individuals based on the probabilities and how to
slice-and-dice these lists. In our campaign, where we want to include
only those people in the targeting range of $[0.3,0.7]$, we would
consider persons 6, 7 and 8. If the campaigns would have plans to
additionally target only those that were younger than 30, then persons
6 and 8 would be targeted.

%s5 #&#
\section{\texorpdfstring{Conclusions.}{Conclusions}}
\label{seccon}
In this paper a framework of statistical methods for targeting for
turnout or targeting for support of eligible voters has been proposed.
It combines ideas of trees with the idea of logistic regression which
was coined LORET. The predictive accuracy performance of different
specifications of LORET estimated with different algorithms has been
investigated for an exemplary data set in a ``targeting for turnout''
setting for a typical situation that a campaign can face: having a
reliable proxy for the target variable at its disposal. Furthermore, we
illustrated how the creation of data-driven voter profiles arises
naturally in the LORET framework and how this can be used for
targeting.

The framework generalizes approaches used by campaigns and is easy to
understand or communicate to people who are familiar with logistic
regression and/or trees. Furthermore, it allows to create a
segmentation of the data which corresponds to automatically building
data-driven voter profiles which can enhance the effectiveness of
targeting measures. As such, the framework is well suited for the
purpose of segmentation and identification in voter
targeting.

Regarding the special cases of LORET, a tree with a logistic node model
may be the most useful default version. For our data, it has the best
cutoff independent predictive accuracy (measured by AUC) and the
highest predictive accuracy (at a cutoff of $0.5$). Please note that in
our study we had completely accurate labels available, so the accuracy
to be expected when dealing only with a proxy from polls might easily
be lower.
The logistic model tree has the additional advantage of providing
refined voter profiles for targeting. As a result, decisions based on
the $y\sim s \Sep e$ LORET are easy to communicate to campaigns that
already use logistic regression or trees.

The other instances of LORET, however, are not without merit either.
Specifically, a LORET of the $y \sim1 \Sep s+e$ type is a good choice
if it is not clear what the functional form in the nodes should look
like or if there is no standard set of variables to be used in the
terminal nodes. Here the nonparametric nature of classification trees
show their advantage. If the targeting situation is such that
the proxies are generally not very reliable/typical for the real
outcome of interest or there is a high degree of noise in these
variables, the extra flexibility and tendency to overfit which trees
exhibit can be a disadvantage. Here, logistic regression may be more
appropriate due to the strict functional relationship that it imposes
and therefore exhibiting less variability in the predictions about the
future. Therefore, even a LORET with just a root node can come in
handy.

We find that if campaigns can use accurate proxy data for the outcome
of interest, the flexibility introduced by the tree structure may lead
to higher predictive accuracy. In a situation where the campaign has to
rely on historic proxy data for the outcome of interest, the predictive
accuracy is generally low and there will probably be little difference
between using a single logistic model or learning partitions as well
(see the rejoinder in the supplementary material [\citet{supplementb}]). We conclude that campaigns are
generally best advised to make an effort in collecting accurate proxies
for the outcome of interest and enabling an analysis as outlined in
Section~\ref{secfcpart}. We believe this is feasible by using
well-designed, repeated polling to obtain the target variable. It is up
to future research to establish what the best proxies to be used as
labels in the targeting stage actually are.

With the benefits mentioned above, one would consider how to
incorporate this technique into the overall campaign strategy. The
primary benefit of using our framework is that campaigns can have
accurate, interpretable, specific individual level identification of
potential voters. This gives campaigns the ability to customize
communications to each individual. Once the campaigns have better
knowledge of the potential voter profiles and the likelihood of them
voting, campaigns can maximize the return for the money spent on
targeting potential voters by communicating on issues that matter to
them and target voters who are likely to be mobilized. The bottom line
here is that the LORET framework does not change the commonly used
campaign tactics but adds a precise and flexible tool that allows to
segment and target the recipients of mobilization messages accurately.
For example, in the decision theoretic framework of optimal campaigning
by \citet{imai2011}, LORET models that employ segmentation could be
used as the building block for estimating heterogenous treatment
effects to yield the posterior distributions of the turnout profiles
[i.e., in steps 1 to 3 of \citet{imai2011}, page~9]. We believe that
by employing the LORET framework, campaigns have a flexible and
versatile toolbox at their disposal that can be customized to meet the
campaign's prevalent requirements and can easily be integrated in the
overall strategy of GOTV targeting.

For further research and practical application, it could be fruitful to
improve aspects of particular interest in GOTV campaigns. For example,
it might be beneficial to use techniques such as artificial neural
networks or ensembles of tree methods to improve predictive accuracy.
Over the course of this study we used random forests, neural networks,
support vector machines, Bayesian additive regression trees and
logistic model trees with boosting to check whether they outperform our
tree models. On our data set their performance was not better than the
performance of the LORET models, so we refrained from investigating
those techniques further and reporting them here (but see the supplementary material [\citet{supplementb}]).
Regularized logistic regression models might prove to be a sensible
alternative to the tree approach, especially in terms of
interpretability and variable selection. Regarding the node models,
semi- or nonparametric models might be of interest as well, especially
when the functional form for the logistic model component is not clear.
For building voter profiles based on a predictive model, mixture models
might also be an interesting alternative.

%sA #&#
\begin{appendix}\label{app}
\section*{\texorpdfstring{Appendix: Computational Details.}{Appendix: Computational Details}}
All calculations have been carried out with the statistical software R
2.12.0--2.15.2 [\citet{rcore11}], using \code{glm()} for logistic
regression. Recursive partitioning infrastructure was provided by the
packages \pkg{party} for \code{mob()} [\citet{zeileis08}] (with \code
{safeGLModel} from
\pkg{mobtools,} \citet{mobtools}] and \code{ctree()} [\citet
{hothorn06}], as well as \pkg{rpart} [\citet{rpart0,rpart}] for CART.
We used the \pkg{ROCR} package [\citet{Sing,rocr}] for calculating and
plotting performance measures and ROC curves and \pkg{multcomp} [\citet
{multcomp}] for the simultaneous confidence intervals.
\end{appendix}

\section*{\texorpdfstring{Acknowledgments.}{Acknowledgments}}
We would like to thank Aristotle, Inc. for lending us their data. We
further thank Chad Gosselink, Allison Lee, Joel Rivlin, Hal Malchow and Brad Chism for valuable
comments on earlier drafts of this paper. We also thank our Editor
Susan Paddock, an anonymous Associate Editor and an anonymous reviewer
for valuable comments and suggestions that helped us to improve the
paper greatly.

\begin{supplement}[id=suppA]
\sname{Supplement A}
\stitle{Data and Code}
\slink[doi]{10.1214/13-AOAS648SUPPA} %[doi,text={...}] - jei reikia suskaldyti doi
\sdatatype{.zip}
\sfilename{aoas648\_suppa.zip}
\sdescription{A~bundle containing the code used to produce the results
of the paper and a snapshot of the data set. Unfortunately we are not
at liberty to share the whole original data set, but were allowed to
include an anonymized, random sample ($N=6544$) of the data.}
\end{supplement}

\begin{supplement}[id=suppB]
\sname{Supplement B}
\stitle{Rejoinder}
\slink[doi]{10.1214/13-AOAS648SUPPB} %[doi,text={...}] - jei reikia suskaldyti doi
\sdatatype{.pdf}
\sfilename{aoas648\_suppb.pdf}
\sdescription{A~rejoinder containing additional analyses of LORET
models with a historic proxy variable and a comparison of LORET models
to high-performance methods like Support Vector Machines, Bayesian
Additive Regression Trees, Artificial Neural Networks, Logistic Model
Trees and Random Forests.}
\end{supplement}

%
% imsref loaded by akundreckaite, 2013-06-27 10:21:06
%

\printaddresses


\begin{thebibliography}{61}
% BibTex style file: ims.bst, 2013-01-28
% Default style options (sort=0,type=number).
% Used options (sort=1,type=nameyear).

%b1 #&#
\bibitem[\protect\citeauthoryear{Albert and Anderson}{1984}]{albert84}
%
\begin{barticle}[mr]
\bauthor{\bsnm{Albert},~\bfnm{A.}\binits{A.}} \AND
\bauthor{\bsnm{Anderson},~\bfnm{J.~A.}\binits{J.~A.}}
(\byear{1984}).
\btitle{On the existence of maximum likelihood estimates in logistic regression
models}.
\bjournal{Biometrika}
\bvolume{71}
\bpages{1--10}.
\bid{doi={10.1093/biomet/71.1.1}, issn={0006-3444}, mr={0738319}}
\bptok{imsref}%
\end{barticle}
%
\endbibitem

%b2 #&#
\bibitem[\protect\citeauthoryear{Arceneaux and
Nickerson}{2009}]{arceneaux2009}
%
\begin{barticle}[author]
\bauthor{\bsnm{Arceneaux},~\bfnm{K.}\binits{K.}} \AND
\bauthor{\bsnm{Nickerson},~\bfnm{D.~W.}\binits{D.~W.}}
(\byear{2009}).
\btitle{Who is mobilized to vote? A re-analysis of 11~field experiments}.
\bjournal{American Journal of Political Science}
\bvolume{53}
\bpages{1--16}.
\bptok{imsref}%
\end{barticle}
%
\endbibitem

%b3 #&#
\bibitem[\protect\citeauthoryear{Blumenthal}{2012}]{Blumenthal2012}
%
\begin{bmisc}[author]
\bauthor{\bsnm{Blumenthal},~\bfnm{M.}\binits{M.}}
(\byear{2012}).
\bhowpublished{Obama campaign polls: How the internal data got it right.
\textit{Huffington Post}. Available at
\texttt{\href{http://www.huffingtonpost.com/2012/11/21/obama-campaign-polls-2012\_n\_2171242.html?utm\_hp\_ref=tw}{http://www.huffingtonpost.com/2012/11/21/obama-}
\href{http://www.huffingtonpost.com/2012/11/21/obama-campaign-polls-2012\_n\_217142.html?utm\_hp\_ref=tw}{campaign-polls-2012\_n\_2171242.html?utm\_hp\_ref=tw}}
[accessed 2012-12-09]}.
\bptok{imsref}%
\end{bmisc}
%
\endbibitem

%b4 #&#
\bibitem[\protect\citeauthoryear{Breiman et~al.}{1984}]{breiman84}
%
\begin{bbook}[author]
\bauthor{\bsnm{Breiman},~\bfnm{L.}\binits{L.}},
\bauthor{\bsnm{Friedman},~\bfnm{J.~H.}\binits{J.~H.}},
\bauthor{\bsnm{Olsen},~\bfnm{R.~A.}\binits{R.~A.}} \AND
\bauthor{\bsnm{Stone},~\bfnm{C.~J.}\binits{C.~J.}}
(\byear{1984}).
\btitle{Classification and Regression Trees}.
\bpublisher{Wadsworth}, \blocation{Pacific Grove, CA}.
\bptok{imsref}%
\end{bbook}
%
\endbibitem

%b5 #&#
\bibitem[\protect\citeauthoryear{Cardy}{2005}]{cardy2005}
%
\begin{barticle}[author]
\bauthor{\bsnm{Cardy},~\bfnm{E.~A.}\binits{E.~A.}}
(\byear{2005}).
\btitle{An experimental field study and persuasion effects of partisan direct
mail and phone calls}.
\bjournal{Annals of the American Academy of Political and Social Science}
\bvolume{601}
\bpages{28--40}.
\bptok{imsref}%
\end{barticle}
%
\endbibitem

%b6 #&#
\bibitem[\protect\citeauthoryear{Chan and Loh}{2004}]{ChanLoh2004}
%
\begin{barticle}[mr]
\bauthor{\bsnm{Chan},~\bfnm{Kin-Yee}\binits{K.-Y.}} \AND
\bauthor{\bsnm{Loh},~\bfnm{Wei-Yin}\binits{W.-Y.}}
(\byear{2004}).
\btitle{L{OTUS}: An algorithm for building accurate and comprehensible logistic
regression trees}.
\bjournal{J. Comput. Graph. Statist.}
\bvolume{13}
\bpages{826--852}.
\bid{doi={10.1198/106186004X13064}, issn={1061-8600}, mr={2109054}}
\bptok{imsref}%
\end{barticle}
%
\endbibitem

%b7 #&#
\bibitem[\protect\citeauthoryear{Chaudhuri et~al.}{1995}]{Chaudhuri1995}
%
\begin{barticle}[mr]
\bauthor{\bsnm{Chaudhuri},~\bfnm{Probal}\binits{P.}},
\bauthor{\bsnm{Lo},~\bfnm{Wen~Da}\binits{W.~D.}},
\bauthor{\bsnm{Loh},~\bfnm{Wei-Yin}\binits{W.-Y.}} \AND
\bauthor{\bsnm{Yang},~\bfnm{Ching~Ching}\binits{C.~C.}}
(\byear{1995}).
\btitle{Generalized regression trees}.
\bjournal{Statist. Sinica}
\bvolume{5}
\bpages{641--666}.
\bid{issn={1017-0405}, mr={1347613}}
\bptok{imsref}%
\end{barticle}
%
\endbibitem

%b8 #&#
\bibitem[\protect\citeauthoryear{Chipman, George and
McCulloch}{2010}]{chipman2010}
%
\begin{barticle}[mr]
\bauthor{\bsnm{Chipman},~\bfnm{Hugh~A.}\binits{H.~A.}},
\bauthor{\bsnm{George},~\bfnm{Edward~I.}\binits{E.~I.}} \AND
\bauthor{\bsnm{McCulloch},~\bfnm{Robert~E.}\binits{R.~E.}}
(\byear{2010}).
\btitle{B{ART}: {B}ayesian additive regression trees}.
\bjournal{Ann. Appl. Stat.}
\bvolume{4}
\bpages{266--298}.
\bid{doi={10.1214/09-AOAS285}, issn={1932-6157}, mr={2758172}}
\bptok{imsref}%
\end{barticle}
%
\endbibitem



%b10 #&#
\bibitem[\protect\citeauthoryear{Cutts and Fieldhouse}{2009}]{cutts2009}
%
\begin{barticle}[author]
\bauthor{\bsnm{Cutts},~\bfnm{D.}\binits{D.}} \AND
\bauthor{\bsnm{Fieldhouse},~\bfnm{E.}\binits{E.}}
(\byear{2009}).
\btitle{What small spatial scales are relevant as electoral contexts for
individual voters? {T}he importance of the household on turnout at the 2001
general election}.
\bjournal{American Journal of Political Science}
\bvolume{53}
\bpages{726--739}.
\bptok{imsref}%
\end{barticle}
%
\endbibitem

%b11 #&#
\bibitem[\protect\citeauthoryear{Denny and Doyle}{2009}]{denny2009}
%
\begin{barticle}[author]
\bauthor{\bsnm{Denny},~\bfnm{K.}\binits{K.}} \AND
\bauthor{\bsnm{Doyle},~\bfnm{O.}\binits{O.}}
(\byear{2009}).
\btitle{Does voting history matter? Analysing persistence in turnout}.
\bjournal{American Journal of Political Science}
\bvolume{53}
\bpages{17--35}.
\bptok{imsref}%
\end{barticle}
%
\endbibitem

%b12 #&#
\bibitem[\protect\citeauthoryear{Finkel}{1993}]{finkel1993}
%
\begin{barticle}[author]
\bauthor{\bsnm{Finkel},~\bfnm{Steven}\binits{S.}}
(\byear{1993}).
\btitle{Reexamining the ``Minimal effects'' model in recent presidential
elections}.
\bjournal{Journal of Politics}
\bvolume{55}
\bpages{1--21}.
\bptok{imsref}%
\end{barticle}
%
\endbibitem

%b13 #&#
\bibitem[\protect\citeauthoryear{Gerber and Green}{2000a}]{gerber2000a}
%
\begin{barticle}[author]
\bauthor{\bsnm{Gerber},~\bfnm{A.~S.}\binits{A.~S.}} \AND
\bauthor{\bsnm{Green},~\bfnm{D.~P.}\binits{D.~P.}}
(\byear{2000}a).
\btitle{The effect of a nonpartisan get-out-the-vote drive: An experimental
study of leafleting}.
\bjournal{Journal of Politics}
\bvolume{62}
\bpages{846--857}.
\bptok{imsref}%
\end{barticle}
%
\endbibitem

%b14 #&#
\bibitem[\protect\citeauthoryear{Gerber and Green}{2000b}]{gerber2000b}
%
\begin{barticle}[author]
\bauthor{\bsnm{Gerber},~\bfnm{Allan~S.}\binits{A.~S.}} \AND
\bauthor{\bsnm{Green},~\bfnm{Donald~P.}\binits{D.~P.}}
(\byear{2000}b).
\btitle{The effects of canvassing, telephone calls, and direct mail on voter
turnout: A field experiment}.
\bjournal{American Political Science Review}
\bvolume{94}
\bpages{656--664}.
\bptok{imsref}%
\end{barticle}
%
\endbibitem

%b15 #&#
\bibitem[\protect\citeauthoryear{Gerber, Green and Green}{2007}]{gerber2003}
%
\begin{barticle}[author]
\bauthor{\bsnm{Gerber},~\bfnm{A.~S.}\binits{A.~S.}},
\bauthor{\bsnm{Green},~\bfnm{D.~P.}\binits{D.~P.}} \AND
\bauthor{\bsnm{Green},~\bfnm{M.}\binits{M.}}
(\byear{2007}).
\btitle{Partisan mail and voter turnout: Results from randomized field
experiments}.
\bjournal{Electoral Studies}
\bvolume{22}
\bpages{563--579}.
\bptok{imsref}%
\end{barticle}
%
\endbibitem

%b16 #&#
\bibitem[\protect\citeauthoryear{Goldstein and Ridout}{2002}]{goldstein2002b}
%
\begin{barticle}[author]
\bauthor{\bsnm{Goldstein},~\bfnm{K.}\binits{K.}} \AND
\bauthor{\bsnm{Ridout},~\bfnm{T.~N.}\binits{T.~N.}}
(\byear{2002}).
\btitle{The politics of participation: Mobilization and turnout over time}.
\bjournal{Political Behavior}
\bvolume{24}
\bpages{3--29}.
\bptok{imsref}%
\end{barticle}
%
\endbibitem

%b17 #&#
\bibitem[\protect\citeauthoryear{Green, Gerber and
Nickerson}{2003}]{green2003a}
%
\begin{barticle}[author]
\bauthor{\bsnm{Green},~\bfnm{D.~P.}\binits{D.~P.}},
\bauthor{\bsnm{Gerber},~\bfnm{A.~S.}\binits{A.~S.}} \AND
\bauthor{\bsnm{Nickerson},~\bfnm{D.~W.}\binits{D.~W.}}
(\byear{2003}).
\btitle{Getting out the vote in local elections: Results from six door-to-door
canvassing experiments}.
\bjournal{Journal of Politics}
\bvolume{65}
\bpages{1083--1096}.
\bptok{imsref}%
\end{barticle}
%
\endbibitem

%b18 #&#
\bibitem[\protect\citeauthoryear{Green and Gerber}{2008}]{green2008}
%
\begin{bbook}[author]
\bauthor{\bsnm{Green},~\bfnm{D.~P.}\binits{D.~P.}} \AND
\bauthor{\bsnm{Gerber},~\bfnm{A.~S.}\binits{A.~S.}}
(\byear{2008}).
\btitle{Get Out the Vote: {H}ow to Increase Voter Turnout},
\bedition{2nd} ed.
\bpublisher{Brookings Institution}, \blocation{Washington DC}.
\bptok{imsref}%
\end{bbook}
%
\endbibitem

%b19 #&#
\bibitem[\protect\citeauthoryear{Green and Kern}{2012}]{green2012}
%
\begin{barticle}[author]
\bauthor{\bsnm{Green},~\bfnm{D.~P.}\binits{D.~P.}} \AND
\bauthor{\bsnm{Kern},~\bfnm{H.~L.}\binits{H.~L.}}
(\byear{2012}).
\btitle{Modeling heterogeneous treatment effects in survey
experiments with
{B}ayesian additive regression trees}.
\bjournal{Public Opinion Quarterly}
\bvolume{76}
\bpages{491--511}.
\bptok{imsref}%
\end{barticle}
%
\endbibitem

%b20 #&#
\bibitem[\protect\citeauthoryear{Hansen and Bowers}{2009}]{hansen2009}
%
\begin{barticle}[mr]
\bauthor{\bsnm{Hansen},~\bfnm{Ben~B.}\binits{B.~B.}} \AND
\bauthor{\bsnm{Bowers},~\bfnm{Jake}\binits{J.}}
(\byear{2009}).
\btitle{Attributing effects to a cluster-randomized get-out-the-vote campaign}.
\bjournal{J. Amer. Statist. Assoc.}
\bvolume{104}
\bpages{873--885}.
\bid{doi={10.1198/jasa.2009.ap06589}, issn={0162-1459}, mr={2562000}}
\bptok{imsref}%
\end{barticle}
%
\endbibitem

%b21 #&#
\bibitem[\protect\citeauthoryear{Hastie, Tibshirani and
Friedman}{2009}]{hastie06}
%
\begin{bbook}[mr]
\bauthor{\bsnm{Hastie},~\bfnm{Trevor}\binits{T.}},
\bauthor{\bsnm{Tibshirani},~\bfnm{Robert}\binits{R.}} \AND
\bauthor{\bsnm{Friedman},~\bfnm{Jerome}\binits{J.}}
(\byear{2009}).
\btitle{The Elements of Statistical Learning: Data Mining, Inference,
and Prediction},
\bedition{2nd} ed.
\bpublisher{Springer}, \blocation{New York}.
\bid{doi={10.1007/978-0-387-84858-7}, mr={2722294}}
\bptok{imsref}%
\end{bbook}
%
\endbibitem

%b22 #&#
\bibitem[\protect\citeauthoryear{Holbrook and McClurg}{2005}]{holbrook2005}
%
\begin{barticle}[author]
\bauthor{\bsnm{Holbrook},~\bfnm{T.~M.}\binits{T.~M.}} \AND
\bauthor{\bsnm{McClurg},~\bfnm{S.~D.}\binits{S.~D.}}
(\byear{2005}).
\btitle{The mobilization of core supporters: Campaigns, turnout and electoral
composition in {U}nited {S}tates presidential elections}.
\bjournal{American Journal of Political Science}
\bvolume{49}
\bpages{689--703}.
\bptok{imsref}%
\end{barticle}
%
\endbibitem

%b23 #&#
\bibitem[\protect\citeauthoryear{Hothorn, Bretz and
Westfall}{2008}]{multcomp}
%
\begin{barticle}[mr]
\bauthor{\bsnm{Hothorn},~\bfnm{Torsten}\binits{T.}},
\bauthor{\bsnm{Bretz},~\bfnm{Frank}\binits{F.}} \AND
\bauthor{\bsnm{Westfall},~\bfnm{Peter}\binits{P.}}
(\byear{2008}).
\btitle{Simultaneous inference in general parametric models}.
\bjournal{Biom. J.}
\bvolume{50}
\bpages{346--363}.
\bid{doi={10.1002/bimj.200810425}, issn={0323-3847}, mr={2521547}}
\bptok{imsref}%
\end{barticle}
%
\endbibitem

%b24 #&#
\bibitem[\protect\citeauthoryear{Hothorn, Hornik and
Zeileis}{2006}]{hothorn06}
%
\begin{barticle}[mr]
\bauthor{\bsnm{Hothorn},~\bfnm{Torsten}\binits{T.}},
\bauthor{\bsnm{Hornik},~\bfnm{Kurt}\binits{K.}} \AND
\bauthor{\bsnm{Zeileis},~\bfnm{Achim}\binits{A.}}
(\byear{2006}).
\btitle{Unbiased recursive partitioning: A~conditional inference framework}.
\bjournal{J. Comput. Graph. Statist.}
\bvolume{15}
\bpages{651--674}.
\bid{doi={10.1198/106186006X133933}, issn={1061-8600}, mr={2291267}}
\bptok{imsref}%
\end{barticle}
%
\endbibitem

%b25 #&#
\bibitem[\protect\citeauthoryear{Hothorn et~al.}{2005}]{hothorn2005}
%
\begin{barticle}[mr]
\bauthor{\bsnm{Hothorn},~\bfnm{Torsten}\binits{T.}},
\bauthor{\bsnm{Leisch},~\bfnm{Friedrich}\binits{F.}},
\bauthor{\bsnm{Zeileis},~\bfnm{Achim}\binits{A.}} \AND
\bauthor{\bsnm{Hornik},~\bfnm{Kurt}\binits{K.}}
(\byear{2005}).
\btitle{The design and analysis of benchmark experiments}.
\bjournal{J. Comput. Graph. Statist.}
\bvolume{14}
\bpages{675--699}.
\bid{doi={10.1198/106186005X59630}, issn={1061-8600}, mr={2170208}}
\bptok{imsref}%
\end{barticle}
%
\endbibitem

%b26 #&#
\bibitem[\protect\citeauthoryear{Imai and Strauss}{2011}]{imai2011}
%
\begin{barticle}[author]
\bauthor{\bsnm{Imai},~\bfnm{K.}\binits{K.}} \AND
\bauthor{\bsnm{Strauss},~\bfnm{A.}\binits{A.}}
(\byear{2011}).
\btitle{Estimation of heterogeneous treatment effects from randomized
experiments, with application to the optimal planning of the get-out-the-vote
campaign}.
\bjournal{Political Analysis}
\bvolume{19}
\bpages{1--19}.
\bptok{imsref}%
\end{barticle}
%
\endbibitem

%b27 #&#
\bibitem[\protect\citeauthoryear{Issenberg}{2012a}]{Issenberg2012}
%
\begin{bmisc}[author]
\bauthor{\bsnm{Issenberg},~\bfnm{Sasha}\binits{S.}}
(\byear{2012}a).
\bhowpublished{Obama's white whale: How the campaign{'}s top-secret
project {N}arwhal
could change this race, and many to come.
\textit{Slate}. Available at
\texttt{\href{http://www.slate.com/articles/news\_and\_politics/victory\_lab/2012/02/project\_narwhal\_how\_a\_top\_secret\_obama\_campaign\_program\_could\_change\_the\_2012\_race\_.html}{http://www.slate.com/articles/news\_and\_politics/victory\_lab/2012/02/project\_}
\href{http://www.slate.com/articles/news\_and\_politics/victory\_lab/2012/02/project\_narwhal\_how\_a\_top\_secret\_obama\_campaign\_program\_could\_change\_the\_2012\_race\_.html}{narwhal\_how\_a\_top\_secret\_obama\_campaign\_program\_could\_change\_the\_2012\_race\_.html}} [accessed 2012-12-04]}.
\bptok{imsref}%
\end{bmisc}
%
\endbibitem

%b28 #&#
\bibitem[\protect\citeauthoryear{Issenberg}{2012b}]{Issenberg2012b}
%
\begin{bbook}[author]
\bauthor{\bsnm{Issenberg},~\bfnm{Sasha}\binits{S.}}
(\byear{2012}b).
\btitle{The Victory Lab: The Secret Science of Winning Campaigns}.
\bpublisher{Crown Publishers}, \blocation{New York}.
\bptok{imsref}%
\end{bbook}
%
\endbibitem

%b29 #&#
\bibitem[\protect\citeauthoryear{Karp and Banducci}{2007}]{karp2007}
%
\begin{barticle}[author]
\bauthor{\bsnm{Karp},~\bfnm{J.~A.}\binits{J.~A.}} \AND
\bauthor{\bsnm{Banducci},~\bfnm{S.~A.}\binits{S.~A.}}
(\byear{2007}).
\btitle{Party mobilization and political participation in new and old
democracies}.
\bjournal{Party Politics}
\bvolume{13}
\bpages{217--234}.
\bptok{imsref}%
\end{barticle}
%
\endbibitem

%b30 #&#
\bibitem[\protect\citeauthoryear{Karp, Banducci and Bowler}{2008}]{karp2008}
%
\begin{barticle}[author]
\bauthor{\bsnm{Karp},~\bfnm{J.~A.}\binits{J.~A.}},
\bauthor{\bsnm{Banducci},~\bfnm{S.~A.}\binits{S.~A.}} \AND
\bauthor{\bsnm{Bowler},~\bfnm{S.}\binits{S.}}
(\byear{2008}).
\btitle{Getting out the vote: Party mobilization in a comparative perspective}.
\bjournal{British Journal of Political Science}
\bvolume{38}
\bpages{91--112}.
\bptok{imsref}%
\end{barticle}
%
\endbibitem

%b31 #&#
\bibitem[\protect\citeauthoryear{Kass}{1980}]{chaid}
%
\begin{barticle}[author]
\bauthor{\bsnm{Kass},~\bfnm{G.~V.}\binits{G.~V.}}
(\byear{1980}).
\btitle{An exploratory technique for investigating large quantities of
categorical data}.
\bjournal{J. R. Stat. Soc. Ser. C. Appl. Stat.}
\bvolume{29}
\bpages{119--127}.
\bptok{imsref}%
\end{barticle}
%
\endbibitem

%b32 #&#
\bibitem[\protect\citeauthoryear{Landwehr, Hall and
Eibe}{2005}]{Landwehr2005}
%
\begin{barticle}[author]
\bauthor{\bsnm{Landwehr},~\bfnm{N.}\binits{N.}},
\bauthor{\bsnm{Hall},~\bfnm{M.}\binits{M.}} \AND
\bauthor{\bsnm{Eibe},~\bfnm{F.}\binits{F.}}
(\byear{2005}).
\btitle{Logistic model trees}.
\bjournal{Machine Learning}
\bvolume{59}
\bpages{161--205}.
\bptok{imsref}%
\end{barticle}
%
\endbibitem

%b33 #&#
\bibitem[\protect\citeauthoryear{Loh and Shih}{1997}]{LohShih1997}
%
\begin{barticle}[mr]
\bauthor{\bsnm{Loh},~\bfnm{Wei-Yin}\binits{W.-Y.}} \AND
\bauthor{\bsnm{Shih},~\bfnm{Yu-Shan}\binits{Y.-S.}}
(\byear{1997}).
\btitle{Split selection methods for classification trees}.
\bjournal{Statist. Sinica}
\bvolume{7}
\bpages{815--840}.
\bid{issn={1017-0405}, mr={1488644}}
\bptok{imsref}%
\end{barticle}
%
\endbibitem

%b34 #&#
\bibitem[\protect\citeauthoryear{Malchow}{2008}]{malchow08}
%
\begin{bbook}[author]
\bauthor{\bsnm{Malchow},~\bfnm{H.}\binits{H.}}
(\byear{2008}).
\btitle{Political Targeting},
\bedition{2nd} ed.
\bpublisher{Predicted Lists, LLC}, \blocation{Sacramento, CA}.
\bptok{imsref}%
\end{bbook}
%
\endbibitem

%b35 #&#
\bibitem[\protect\citeauthoryear{McCullagh and Nelder}{1989}]{mccullough89}
%
\begin{bbook}[author]
\bauthor{\bsnm{McCullagh},~\bfnm{Peter}\binits{P.}} \AND
\bauthor{\bsnm{Nelder},~\bfnm{John~A.}\binits{J.~A.}}
(\byear{1989}).
\btitle{Generalized Linear Models},
\bedition{2nd} ed.
\bpublisher{Chapman \& Hall}, \blocation{New York}.
\bptok{imsref}%
\end{bbook}
%
\endbibitem

%b36 #&#
\bibitem[\protect\citeauthoryear{McDonald}{2012}]{mcdonald2012}
%
\begin{bmisc}[author]
\bauthor{\bsnm{McDonald},~\bfnm{Michael}\binits{M.}}
(\byear{2012}).
\bhowpublished{Turnout rates, 1980--2010.
{U}nited {S}tates Election Project.
Available at \url{http://elections.gmu.edu/} [accessed 2012-02-16].}
\bptok{imsref}%
\end{bmisc}
%
\endbibitem

%b37 #&#
\bibitem[\protect\citeauthoryear{Muller}{1999}]{muller1999}
%
\begin{barticle}[author]
\bauthor{\bsnm{Muller},~\bfnm{M.~G.}\binits{M.~G.}}
(\byear{1999}).
\btitle{Electoral campaigning as an occupation---The
professionalization of
political consultants in the {U}nited {S}tates}.
\bjournal{Politische Vierteljahresschrift}
\bvolume{40}
\bpages{198--199}.
\bptok{imsref}%
\end{barticle}
%
\endbibitem

%b38 #&#
\bibitem[\protect\citeauthoryear{Murray and Scime}{2010}]{murray2010}
%
\begin{barticle}[author]
\bauthor{\bsnm{Murray},~\bfnm{Gregg~R.}\binits{G.~R.}} \AND
\bauthor{\bsnm{Scime},~\bfnm{Anthony}\binits{A.}}
(\byear{2010}).
\btitle{Microtargeting and electorate segmentation: Data mining the American
national election studies}.
\bjournal{Journal of Political Marketing}
\bvolume{9}
\bpages{143--166}.
\bptok{imsref}%
\end{barticle}\
%
\endbibitem

%b39 #&#
\bibitem[\protect\citeauthoryear{Nickerson, Friedrichs and
King}{2006}]{nickerson2006}
%
\begin{barticle}[author]
\bauthor{\bsnm{Nickerson},~\bfnm{D.~W.}\binits{D.~W.}},
\bauthor{\bsnm{Friedrichs},~\bfnm{R.~D.}\binits{R.~D.}} \AND
\bauthor{\bsnm{King},~\bfnm{D.~C.}\binits{D.~C.}}
(\byear{2006}).
\btitle{Partisan mobilization campaigns in the field: Results from a statewide
turnout experiment in {M}ichigan}.
\bjournal{Political Research Quarterly}
\bvolume{59}
\bpages{85--97}.
\bptok{imsref}%
\end{barticle}
%
\endbibitem

%b40 #&#
\bibitem[\protect\citeauthoryear{Panagopoulos}{2009}]{panagopoulos2009b}
%
\begin{barticle}[author]
\bauthor{\bsnm{Panagopoulos},~\bfnm{C.}\binits{C.}}
(\byear{2009}).
\btitle{Partisan and nonpartisan message content and voter
mobilization field
experimental evidence}.
\bjournal{Political Research Quarterly}
\bvolume{62}
\bpages{70--76}.
\bptok{imsref}%
\end{barticle}
%
\endbibitem

%b41 #&#
\bibitem[\protect\citeauthoryear{Parry et~al.}{2008}]{parry2008}
%
\begin{barticle}[author]
\bauthor{\bsnm{Parry},~\bfnm{J.}\binits{J.}},
\bauthor{\bsnm{Barth},~\bfnm{J.}\binits{J.}},
\bauthor{\bsnm{Kropf},~\bfnm{M.}\binits{M.}} \AND
\bauthor{\bsnm{Jones},~\bfnm{E.~T.}\binits{E.~T.}}
(\byear{2008}).
\btitle{Mobilizing the seldom voter: Campaign contact and effects in
high-profile elections}.
\bjournal{Political Behavior}
\bvolume{30}
\bpages{97--113}.
\bptok{imsref}%
\end{barticle}
%
\endbibitem

%b42 #&#
\bibitem[\protect\citeauthoryear{Phillips, Urbany and
Reynolds}{2008}]{philips2008}
%
\begin{barticle}[author]
\bauthor{\bsnm{Phillips},~\bfnm{J.~M.}\binits{J.~M.}},
\bauthor{\bsnm{Urbany},~\bfnm{J.~E.}\binits{J.~E.}} \AND
\bauthor{\bsnm{Reynolds},~\bfnm{T.~J.}\binits{T.~J.}}
(\byear{2008}).
\btitle{Confirmation and the effects of valenced political
advertising: A field
experiment}.
\bjournal{Journal of Consumer Research}
\bvolume{34}
\bpages{794--806}.
\bptok{imsref}%
\end{barticle}
%
\endbibitem

%b43 #&#
\bibitem[\protect\citeauthoryear{Plasser}{2000}]{plasser2000}
%
\begin{barticle}[author]
\bauthor{\bsnm{Plasser},~\bfnm{F.}\binits{F.}}
(\byear{2000}).
\btitle{{A}merican campaign techniques worldwide}.
\bjournal{Harvard International Journal of Press-Politics}
\bvolume{5}
\bpages{33--54}.
\bptok{imsref}%
\end{barticle}
%
\endbibitem

%b44 #&#
\bibitem[\protect\citeauthoryear{Quinlan}{1993}]{Quinlan1993}
%
\begin{bbook}[author]
\bauthor{\bsnm{Quinlan},~\bfnm{J.~R.}\binits{J.~R.}}
(\byear{1993}).
\btitle{C4.5: Programs for Machine Learning}.
\bpublisher{Morgan Kaufmann}, \blocation{San Mateo, CA}.
\bptok{imsref}%
\end{bbook}
%
\endbibitem

%b52 #&#
\bibitem[\protect\citeauthoryear{{R Development Core Team}}{2012}]{rcore11}
%
\begin{bmisc}[author]
\borganization{R Development Core Team}
(\byear{2012}).
\bhowpublished{{R}: A language and environment for statistical computing.
R Foundation for Statstical Computing, Vienna, Austria}.
\bptok{imsref}%
\end{bmisc}
%
\endbibitem

%b45 #&#
\bibitem[\protect\citeauthoryear{Rusch and Zeileis}{2013}]{Rusch2012}
%
\begin{barticle}[author]
\bauthor{\bsnm{Rusch},~\bfnm{Thomas}\binits{T.}} \AND
\bauthor{\bsnm{Zeileis},~\bfnm{Achim}\binits{A.}}
(\byear{2013}).
\btitle{Gaining insight with recursive partitioning of
generalized linear
models.}
\bjournal{J. Stat. Comput. Simul.}
\bvolume{83}
\bpages{1301--1315}.
\bptok{imsref}%
\end{barticle}
%
\endbibitem

%b46 #&#
\bibitem[\protect\citeauthoryear{Rusch et~al.}{2012}]{mobtools}
%
\begin{bmisc}[author]
\bauthor{\bsnm{Rusch},~\bfnm{T.}\binits{T.}},
\bauthor{\bsnm{Zeileis},~\bfnm{A.}\binits{A.}},
\bauthor{\bsnm{Hothorn},~\bfnm{T.}\binits{T.}} \AND
\bauthor{\bsnm{Leisch},~\bfnm{F.}\binits{F.}}
(\byear{2012}).
\bhowpublished{mobtools: A collection of {statmodel}s and of utilities
for extending
{mob}. {R} package version 0.0-1}.
\bptok{imsref}%
\end{bmisc}
%
\endbibitem

%b47 #&#
\bibitem[\protect\citeauthoryear{Rusch et~al.}{2013a}]{supplementa}
%
\begin{bmisc}[author]
\bauthor{\bsnm{Rusch},~\bfnm{Thomas}\binits{T.}},
\bauthor{\bsnm{Lee},~\bfnm{Ilro}\binits{I.}},
\bauthor{\bsnm{Hornik},~\bfnm{Kurt}\binits{K.}},
\bauthor{\bsnm{Jank},~\bfnm{Wolfgang}\binits{W.}} \AND
\bauthor{\bsnm{Zeileis},~\bfnm{Achim}\binits{A.}}
(\byear{2013a}).
\bhowpublished{Supplement to ``Influencing elections with
statistics:
Targeting voters with logistic regression trees.'' DOI:\doiurl
{10.1214/13-AOAS648SUPPA}}.
\bptok{imsref}%
\end{bmisc}
%
\endbibitem


%b47 #&#
\bibitem[\protect\citeauthoryear{Rusch et~al.}{2013b}]{supplementb}
%
\begin{bmisc}[author]
\bauthor{\bsnm{Rusch},~\bfnm{Thomas}\binits{T.}},
\bauthor{\bsnm{Lee},~\bfnm{Ilro}\binits{I.}},
\bauthor{\bsnm{Hornik},~\bfnm{Kurt}\binits{K.}},
\bauthor{\bsnm{Jank},~\bfnm{Wolfgang}\binits{W.}} \AND
\bauthor{\bsnm{Zeileis},~\bfnm{Achim}\binits{A.}}
(\byear{2013b}).
\bhowpublished{Supplement to ``Influencing elections with
statistics:
Targeting voters with logistic regression trees.'' DOI:\doiurl
{10.1214/13-AOAS648SUPPB}}.
\bptok{imsref}%
\end{bmisc}
%
\endbibitem

%b48 #&#
\bibitem[\protect\citeauthoryear{Sing et~al.}{2005}]{Sing}
%
\begin{barticle}[pbm]
\bauthor{\bsnm{Sing},~\bfnm{Tobias}\binits{T.}},
\bauthor{\bsnm{Sander},~\bfnm{Oliver}\binits{O.}},
\bauthor{\bsnm{Beerenwinkel},~\bfnm{Niko}\binits{N.}} \AND
\bauthor{\bsnm{Lengauer},~\bfnm{Thomas}\binits{T.}}
(\byear{2005}).
\btitle{ROCR: Visualizing classifier performance in R}.
\bjournal{Bioinformatics}
\bvolume{21}
\bpages{3940--3941}.
\bid{doi={10.1093/bioinformatics/bti623}, issn={1367-4803}, pii={bti623},
pmid={16096348}}
\bptok{imsref}%
\end{barticle}
%
\endbibitem

%b49 #&#
\bibitem[\protect\citeauthoryear{Sing et~al.}{2009}]{rocr}
%
\begin{bmisc}[author]
\bauthor{\bsnm{Sing},~\bfnm{Tobias}\binits{T.}},
\bauthor{\bsnm{Sander},~\bfnm{Oliver}\binits{O.}},
\bauthor{\bsnm{Beerenwinkel},~\bfnm{Niko}\binits{N.}} \AND
\bauthor{\bsnm{Lengauer},~\bfnm{Thomas}\binits{T.}}
(\byear{2009}).
\bhowpublished{{ROCR}: Visualizing the performance of scoring
classifiers.
R~package version~1.0-4}.
\bptok{imsref}%
\end{bmisc}
%
\endbibitem

%b50 #&#
\bibitem[\protect\citeauthoryear{Susan}{1999}]{Susan1999}
%
\begin{barticle}[author]
\bauthor{\bsnm{Susan},~\bfnm{J.~C.}\binits{J.~C.}}
(\byear{1999}).
\btitle{The disempowerment of the gender gap: Soccer moms and the 1996
elections}.
\bjournal{PS: Political Science \& Politics}
\bvolume{32}
\bpages{7--11}.
\bptok{imsref}%
\end{barticle}
%
\endbibitem

%b51 #&#
\bibitem[\protect\citeauthoryear{Sussman and Galizio}{2003}]{sussman2003}
%
\begin{barticle}[author]
\bauthor{\bsnm{Sussman},~\bfnm{G.}\binits{G.}} \AND
\bauthor{\bsnm{Galizio},~\bfnm{L.}\binits{L.}}
(\byear{2003}).
\btitle{The global reproduction of {A}merican politics}.
\bjournal{Political Comunication}
\bvolume{20}
\bpages{309--328}.
\bptok{imsref}%
\end{barticle}
%
\endbibitem



%b53 #&#
\bibitem[\protect\citeauthoryear{Therneau and Atkinson}{1997}]{rpart0}
%
\begin{bmisc}[author]
\bauthor{\bsnm{Therneau},~\bfnm{T.~M.}\binits{T.~M.}} \AND
\bauthor{\bsnm{Atkinson},~\bfnm{E.~J.}\binits{E.~J.}}
(\byear{1997}).
\bhowpublished{An introduction to recursive partitioning using the {rpart}
routine. Technical Report {61}, {Section of Biostatistics, Mayo Clinic,
Rochester, NY}.}
\bptok{imsref}%
\end{bmisc}
%
\endbibitem

%b54 #&#
\bibitem[\protect\citeauthoryear{Therneau, Atkinson and Ripley}{2011}]{rpart}
%
\begin{bmisc}[author]
\bauthor{\bsnm{Therneau},~\bfnm{T.~M.}\binits{T.~M.}},
\bauthor{\bsnm{Atkinson},~\bfnm{E.~J.}\binits{E.~J.}} \AND
\bauthor{\bsnm{Ripley},~\bfnm{Brian~D.}\binits{B.~D.}}
(\byear{2011}).
\bhowpublished{rpart: Recursive partitioning.
R~package version 3.1-50}.
\bptok{imsref}%
\end{bmisc}
%
\endbibitem

%b9 #&#
\bibitem[\protect\citeauthoryear{{US Election Assistance
Commission}}{2010}]{congress}
%
\begin{bmisc}[author]
\borganization{US Election Assistance Commission}
(\byear{2010}).
\bhowpublished{The Impact of the {N}ational {V}oter {R}egistration
{A}ct of 1993 on
the Administration of Elections for {F}ederal {O}ffice 2009--2010}.
\bptok{imsref}%
\end{bmisc}
%
\endbibitem

%b55 #&#
\bibitem[\protect\citeauthoryear{Whitelock, Whitelock and van
Heerde}{2010}]{whitelock2010}
%
\begin{barticle}[author]
\bauthor{\bsnm{Whitelock},~\bfnm{A.}\binits{A.}},
\bauthor{\bsnm{Whitelock},~\bfnm{J.}\binits{J.}} \AND
\bauthor{\bparticle{van} \bsnm{Heerde},~\bfnm{J.}\binits{J.}}
(\byear{2010}).
\btitle{The influence of promotional activity and different electoral systems
on voter turnout: A study of the {UK} and {G}erman {E}uro elections}.
\bjournal{European Journal of Marketing}
\bvolume{44}
\bpages{401--420}.
\bptok{imsref}%
\end{barticle}
%
\endbibitem

%b56 #&#
\bibitem[\protect\citeauthoryear{Wielhouwer}{2003}]{wielhouwer2003}
%
\begin{barticle}[author]
\bauthor{\bsnm{Wielhouwer},~\bfnm{Peter~W.}\binits{P.~W.}}
(\byear{2003}).
\btitle{In search of Lincoln's perfect list---Targeting in grassroots
campaigns}.
\bjournal{American Politics Research}
\bvolume{31}
\bpages{632--669}.
\bptok{imsref}%
\end{barticle}
%
\endbibitem

%b57 #&#
\bibitem[\protect\citeauthoryear{Wikipedia}{2012}]{Wikielection}
%
\begin{bmisc}[author]
\borganization{Wikipedia}
(\byear{2012}).
\bhowpublished{{U}nited {S}tates presidential election, 2012.
Available at \texttt{\href{http://en.wikipedia.org/wiki/Unite\_States\_presidential\_election,\_2012}{http://en.}
\href{http://en.wikipedia.org/wiki/Unite\_States\_presidential\_election,\_2012}{wikipedia.org/wiki/Unite\_States\_presidential\_election,\_2012}} [accessed 2012-11-21]}.
\bptok{imsref}%
\end{bmisc}
%
\endbibitem

%b58 #&#
\bibitem[\protect\citeauthoryear{Wilcoxon}{1945}]{Wilcoxon1945}
%
\begin{barticle}[author]
\bauthor{\bsnm{Wilcoxon},~\bfnm{Frank}\binits{F.}}
(\byear{1945}).
\btitle{Individual comparisons by ranking methods}.
\bjournal{Biometrics Bulletin}
\bvolume{1}
\bpages{80--83}.
\bptok{imsref}%
\end{barticle}
%
\endbibitem

%b59 #&#
\bibitem[\protect\citeauthoryear{Zeileis, Hothorn and
Hornik}{2008}]{zeileis08}
%
\begin{barticle}[mr]
\bauthor{\bsnm{Zeileis},~\bfnm{Achim}\binits{A.}},
\bauthor{\bsnm{Hothorn},~\bfnm{Torsten}\binits{T.}} \AND
\bauthor{\bsnm{Hornik},~\bfnm{Kurt}\binits{K.}}
(\byear{2008}).
\btitle{Model-based recursive partitioning}.
\bjournal{J. Comput. Graph. Statist.}
\bvolume{17}
\bpages{492--514}.
\bid{doi={10.1198/106186008X319331}, issn={1061-8600}, mr={2439970}}
\bptok{imsref}%
\end{barticle}
%
\endbibitem

%b60 #&#
\bibitem[\protect\citeauthoryear{Zhang and Singer}{2010}]{Zhang2010}
%
\begin{bbook}[mr]
\bauthor{\bsnm{Zhang},~\bfnm{Heping}\binits{H.}} \AND
\bauthor{\bsnm{Singer},~\bfnm{Burton~H.}\binits{B.~H.}}
(\byear{2010}).
\btitle{Recursive Partitioning and Applications},
\bedition{2nd} ed.
\bpublisher{Springer}, \blocation{New York}.
\bid{doi={10.1007/978-1-4419-6824-1}, mr={2674991}}
\bptok{imsref}%
\end{bbook}
%
\endbibitem

\end{thebibliography}
\end{document}